\documentclass[reqno,12pt,a4paper]{amsart}
\usepackage{mathrsfs}
\usepackage{amssymb}
\usepackage{amsfonts}
\usepackage{latexsym}
\usepackage{amsthm}

\usepackage{graphicx}
\def\lb{\label}

\newcommand{\er}[1]{\textrm{(\ref{#1})}}

\begin{document}

\renewcommand{\theequation}{\arabic{section}.\arabic{equation}}
\theoremstyle{plain}
\newtheorem{theorem}{\bf Theorem}[section]
\newtheorem{lemma}[theorem]{\bf Lemma}
\newtheorem{corollary}[theorem]{\bf Corollary}
\newtheorem{proposition}[theorem]{\bf Proposition}
\newtheorem{definition}[theorem]{\bf Definition}
\newtheorem{condition}[theorem]{\bf Condition}
\newtheorem{remark}[theorem]{\it Remark}

\def\a{\alpha}  \def\cA{{\mathcal A}}     \def\bA{{\bf A}}  \def\mA{{\mathscr A}}
\def\b{\beta}   \def\cB{{\mathcal B}}     \def\bB{{\bf B}}  \def\mB{{\mathscr B}}
\def\g{\gamma}  \def\cC{{\mathcal C}}     \def\bC{{\bf C}}  \def\mC{{\mathscr C}}
\def\G{\Gamma}  \def\cD{{\mathcal D}}     \def\bD{{\bf D}}  \def\mD{{\mathscr D}}
\def\d{\delta}  \def\cE{{\mathcal E}}     \def\bE{{\bf E}}  \def\mE{{\mathscr E}}
\def\D{\Delta}  \def\cF{{\mathcal F}}     \def\bF{{\bf F}}  \def\mF{{\mathscr F}}
\def\c{\chi}    \def\cG{{\mathcal G}}     \def\bG{{\bf G}}  \def\mG{{\mathscr G}}
\def\z{\zeta}   \def\cH{{\mathcal H}}     \def\bH{{\bf H}}  \def\mH{{\mathscr H}}
\def\e{\eta}    \def\cI{{\mathcal I}}     \def\bI{{\bf I}}  \def\mI{{\mathscr I}}
\def\p{\psi}    \def\cJ{{\mathcal J}}     \def\bJ{{\bf J}}  \def\mJ{{\mathscr J}}
\def\vT{\Theta} \def\cK{{\mathcal K}}     \def\bK{{\bf K}}  \def\mK{{\mathscr K}}
\def\k{\kappa}  \def\cL{{\mathcal L}}     \def\bL{{\bf L}}  \def\mL{{\mathscr L}}
\def\l{\lambda} \def\cM{{\mathcal M}}     \def\bM{{\bf M}}  \def\mM{{\mathscr M}}
\def\L{\Lambda} \def\cN{{\mathcal N}}     \def\bN{{\bf N}}  \def\mN{{\mathscr N}}
\def\m{\mu}     \def\cO{{\mathcal O}}     \def\bO{{\bf O}}  \def\mO{{\mathscr O}}
\def\n{\nu}     \def\cP{{\mathcal P}}     \def\bP{{\bf P}}  \def\mP{{\mathscr P}}
\def\r{\rho}    \def\cQ{{\mathcal Q}}     \def\bQ{{\bf Q}}  \def\mQ{{\mathscr Q}}
\def\s{\sigma}  \def\cR{{\mathcal R}}     \def\bR{{\bf R}}  \def\mR{{\mathscr R}}
\def\S{\Sigma}  \def\cS{{\mathcal S}}     \def\bS{{\bf S}}  \def\mS{{\mathscr S}}
\def\t{\tau}    \def\cT{{\mathcal T}}     \def\bT{{\bf T}}  \def\mT{{\mathscr T}}
\def\f{\phi}    \def\cU{{\mathcal U}}     \def\bU{{\bf U}}  \def\mU{{\mathscr U}}
\def\F{\Phi}    \def\cV{{\mathcal V}}     \def\bV{{\bf V}}  \def\mV{{\mathscr V}}
\def\P{\Psi}    \def\cW{{\mathcal W}}     \def\bW{{\bf W}}  \def\mW{{\mathscr W}}
\def\o{\omega}  \def\cX{{\mathcal X}}     \def\bX{{\bf X}}  \def\mX{{\mathscr X}}
\def\x{\xi}     \def\cY{{\mathcal Y}}     \def\bY{{\bf Y}}  \def\mY{{\mathscr Y}}
\def\X{\Xi}     \def\cZ{{\mathcal Z}}     \def\bZ{{\bf Z}}  \def\mZ{{\mathscr Z}}
\def\O{\Omega}
\def\th{\theta}

\newcommand{\gA}{\mathfrak{A}}
\newcommand{\gB}{\mathfrak{B}}
\newcommand{\gC}{\mathfrak{C}}
\newcommand{\gD}{\mathfrak{D}}
\newcommand{\gE}{\mathfrak{E}}
\newcommand{\gF}{\mathfrak{F}}
\newcommand{\gG}{\mathfrak{G}}
\newcommand{\gH}{\mathfrak{H}}
\newcommand{\gI}{\mathfrak{I}}
\newcommand{\gJ}{\mathfrak{J}}
\newcommand{\gK}{\mathfrak{K}}
\newcommand{\gL}{\mathfrak{L}}
\newcommand{\gM}{\mathfrak{M}}
\newcommand{\gN}{\mathfrak{N}}
\newcommand{\gO}{\mathfrak{O}}
\newcommand{\gP}{\mathfrak{P}}
\newcommand{\gQ}{\mathfrak{Q}}
\newcommand{\gR}{\mathfrak{R}}
\newcommand{\gS}{\mathfrak{S}}
\newcommand{\gT}{\mathfrak{T}}
\newcommand{\gU}{\mathfrak{U}}
\newcommand{\gV}{\mathfrak{V}}
\newcommand{\gW}{\mathfrak{W}}
\newcommand{\gX}{\mathfrak{X}}
\newcommand{\gY}{\mathfrak{Y}}
\newcommand{\gZ}{\mathfrak{Z}}

\newcommand{\gm}{\mathfrak{m}}
\newcommand{\gn}{\mathfrak{n}}
\newcommand{\gf}{\mathfrak{f}}
\newcommand{\gh}{\mathfrak{h}}
\newcommand{\mg}{\mathfrak{g}}
\newcommand{\gb}{\mathfrak{b}}
\newcommand{\ga}{\mathfrak{a}}
\newcommand{\gu}{\mathfrak{u}}
\newcommand{\gv}{\mathfrak{v}}
\newcommand{\gw}{\mathfrak{w}}

\def\ve{\varepsilon}   \def\vt{\vartheta}    \def\vp{\varphi}    \def\vk{\varkappa}

\def\Z{{\mathbb Z}}    \def\R{{\mathbb R}}   \def\C{{\mathbb C}}    \def\K{{\mathbb K}}
\def\T{{\mathbb T}}    \def\N{{\mathbb N}}   \def\dD{{\mathbb D}}


\def\la{\leftarrow}              \def\ra{\rightarrow}            \def\Ra{\Rightarrow}
\def\ua{\uparrow}                \def\da{\downarrow}
\def\lra{\leftrightarrow}        \def\Lra{\Leftrightarrow}


\def\lt{\biggl}                  \def\rt{\biggr}
\def\ol{\overline}               \def\wt{\widetilde}
\def\no{\noindent}


\let\ge\geqslant                 \let\le\leqslant
\def\lan{\langle}                \def\ran{\rangle}
\def\/{\over}                    \def\iy{\infty}
\def\sm{\setminus}               \def\es{\emptyset}
\def\ss{\subset}                 \def\ts{\times}
\def\pa{\partial}                \def\os{\oplus}
\def\om{\ominus}                 \def\ev{\equiv}
\def\iint{\int\!\!\!\int}        \def\iintt{\mathop{\int\!\!\int\!\!\dots\!\!\int}\limits}
\def\el2{\ell^{\,2}}             \def\1{1\!\!1}
\def\sh{\sharp}
\def\wh{\widehat}
\def\bs{\backslash}

\def\sh{\mathop{\mathrm{sh}}\nolimits}
\def\Area{\mathop{\mathrm{Area}}\nolimits}
\def\arg{\mathop{\mathrm{arg}}\nolimits}
\def\const{\mathop{\mathrm{const}}\nolimits}
\def\det{\mathop{\mathrm{det}}\nolimits}
\def\diag{\mathop{\mathrm{diag}}\nolimits}
\def\diam{\mathop{\mathrm{diam}}\nolimits}
\def\dim{\mathop{\mathrm{dim}}\nolimits}
\def\dist{\mathop{\mathrm{dist}}\nolimits}
\def\Im{\mathop{\mathrm{Im}}\nolimits}
\def\Iso{\mathop{\mathrm{Iso}}\nolimits}
\def\Ker{\mathop{\mathrm{Ker}}\nolimits}
\def\Lip{\mathop{\mathrm{Lip}}\nolimits}
\def\rank{\mathop{\mathrm{rank}}\limits}
\def\Ran{\mathop{\mathrm{Ran}}\nolimits}
\def\Re{\mathop{\mathrm{Re}}\nolimits}
\def\Res{\mathop{\mathrm{Res}}\nolimits}
\def\res{\mathop{\mathrm{res}}\limits}
\def\sign{\mathop{\mathrm{sign}}\nolimits}
\def\span{\mathop{\mathrm{span}}\nolimits}
\def\supp{\mathop{\mathrm{supp}}\nolimits}
\def\Tr{\mathop{\mathrm{Tr}}\nolimits}
\def\BBox{\hspace{1mm}\vrule height6pt width5.5pt depth0pt \hspace{6pt}}
\def\as{\text{as}}
\def\all{\text{all}}
\def\where{\text{where}}
\def\Dom{\mathop{\mathrm{Dom}}\nolimits}


\newcommand\nh[2]{\widehat{#1}\vphantom{#1}^{(#2)}}
\def\dia{\diamond}

\def\Oplus{\bigoplus\nolimits}



\def\qqq{\qquad}
\def\qq{\quad}
\let\ge\geqslant
\let\le\leqslant
\let\geq\geqslant
\let\leq\leqslant
\newcommand{\ca}{\begin{cases}}
\newcommand{\ac}{\end{cases}}
\newcommand{\ma}{\begin{pmatrix}}
\newcommand{\am}{\end{pmatrix}}
\renewcommand{\[}{\begin{equation}}
\renewcommand{\]}{\end{equation}}
\def\eq{\begin{equation}}
\def\qe{\end{equation}}
\def\[{\begin{equation}}
\def\bu{\bullet}

\newcommand{\fr}{\frac}
\newcommand{\tf}{\tfrac}

\title[Asymptotics of the divisor for the good Boussinesq equation]
{Asymptotics of the divisor for the good Boussinesq equation}

\date{\today}
\author[Andrey Badanin]{Andrey Badanin}
\author[Evgeny Korotyaev]{Evgeny L. Korotyaev}
\address{Saint-Petersburg
State University, Universitetskaya nab. 7/9, St. Petersburg,
199034 Russia,
a.badanin@spbu.ru,\
korotyaev@gmail.com,\  e.korotyaev@spbu.ru}


\maketitle

\begin{abstract}
We consider a third order operator under the three-point Dirichlet condition.
Its  spectrum is the so-called auxiliary spectrum
for the good Boussinesq equation,
as well as the Dirichlet spectrum
for the Schr\"odinger operator on the unit interval is the auxiliary spectrum for the periodic KdV equation.
The auxiliary spectrum is formed by projections
of the points of the divisor onto the spectral plane.
We estimate the spectrum and the corresponding norming
constants in terms of small operator coefficients. This work is the first in a series of papers
devoted to solving the inverse problem for the Boussinesq equation.
\end{abstract}

\section{Introduction and main results}
\setcounter{equation}{0}

\subsection{Introduction}
The operator $\pa^3+p\pa+\pa p+q,\pa=\pa/\pa x$,
with periodic $p$ and $q$, is  {\it the Lax operator}
for the periodic good Boussinesq equation (below the GBE)
$$
p_{tt}=-{1\/3}(p_{xxxx}+4(p^2)_{xx}), \qqq p_t=q_x,
$$
see \cite{McK81}. In order to solve the GBE, we need spectral
analysis of this operator. Solutions of the the GBE can be described
in terms of two objects. The first one is {\it the  Riemann surface}
of the Lax operator. It is invariant with respect to the Boussinesq
flow. The second one is a set of points on a Riemann surface, {\it a
divisor}. It parameterizes solutions to the Boussinesq equation. The
projections of the points of the divisor onto the spectral plane
form {\it the auxiliary spectrum} of the Lax operator.

We discuss our plan for studying the inverse spectral problem for
the Boussinesq equation on a circle. The experience with the Hill
operator \cite{K99} shows us that we need three sets of spectral
data.

 1. The set of branch points of the Riemann surface for the operator,
this set coincides with the spectrum of the 2-periodic problem,

 2. The auxiliary spectrum, which coincides with the spectrum
of the Dirichlet problem.

 3. The sequence  $(s_n)_{n\in\N}$ of signs $s_n=+$ or  $s_n=-$, i.e.,
 the sign of the norming constants.

\no  The branch points
of the Riemann surface are the edges of the gaps in the spectrum
of the operator on the line.
Each finite gap contains exactly one eigenvalue of the Dirichlet
problem. Moreover, when the potential is shifted by a period, the
eigenvalue of the Dirichlet problem, which lies in the n-th gap,
runs through it monotonically from one edge to the other, making
exactly n revolutions. Note that in \cite{MO75}
(see also \cite{K97}, where the more simple proof is given) the authors work
with the Riemann surface in terms of the quasimomentum, which is
more complicated.

The situation for the Boussinesq equation is similar
to that for the KdV equation, see \cite{BK24xxx}.
In order to uniquely define
the coefficients $p,q$ we need 3 sets of the spectral data.

\no 1. The set of branch points of the Riemann surface for the Lax operator.
The branch points, with exception of a finite number, are real, while
the 2-periodic spectrum asymptotically lies near the line parallel to the
imaginary axis.

\no  2. The auxiliary spectrum coincides with the spectrum
of the 3-point Dirichlet problem.

\no 3. The  sequence  $(s_n)_{n\in\Z\sm\{0\}}$ of signs $s_n=+$ or  $s_n=-$
 , i.e.,  the sign of the norming constants.

\no In our case of a 3rd order operator,
the set of branch points is not the spectrum of any problem associated
with the operator. But it is still real, with the possible exception
of a finite number of points. There are some intervals on the Riemann surface,
``pseudo-gaps'', similar to the gaps in the spectrum of the Hill operator.
Edges of these intervals are adjacent branch points.
The 3-point Dirichlet eigenvalues, with the exception
of a finite number, are real, each ``pseudo-gap'' contains exactly
one eigenvalue, and when the potential is shifted by a period,
the eigenvalue of the problem lying in the n-th ``pseudo-gap''
runs monotonically
from one edge to the other, making exactly n turns.

Moreover, there is an important relation between
the spectral data for the 3rd order operator and
the spectral data for the Hill operator.
In fact, McKean \cite{McK81} introduced a transformation of
the 3rd order equation
with 1-periodic coefficients
to the Hill equation with an  energy-dependent
potential. He showed that under this transformation the set of
ramifications turns into a set of eigenvalues
of the 2-periodic problem for the Schr\"odinger operator.
Additionally, in \cite{BK24x} we show that under
McKean's transformation the set of the
eigenvalues of the 3-point Dirichlet problem for the 3-order
operator turns into the set of the eigenvalues of the Dirichlet problem
for the Schr\"odinger operator.

In general, some eigenvalues of the 3-point problem may be non-real
and have multiplicity $\ge 2$. However,
if $p$ and $q$ are small, then all eigenvalues are real and simple.
In \cite{BK24xx} we construct the mapping from the space of coefficients
$p,q$ onto the space of spectral data and prove, that this mapping is
a real analytic bijection between the neighborhood of the point $p=q=0$
onto the image of this neighborhood. Ultimately, this leads to the
fact that a solution of the Boussinesq equation with sufficiently
small initial data from $L^2(\T)$ exists globally
in this class. In the case when the coefficients of the operator are
not small, the eigenvalues can be non-real and multiples. This greatly
complicates the analysis and leads to the fact that the solutions of
the Boussinesq equation have a blow-up discovered by Kalantarov and
Ladyzhenskaja \cite{KL77}. The corresponding inverse problem still needs
to be solved.

Our paper is devoted to the auxiliary spectrum and the corresponding norming constants for the Boussinesq
equation. We consider a non-self-adjoint operator $\cL=\cL(\gu)$
acting on $L^2(0,2)$ and given by
\[
\lb{Hdpq}
\cL y=(y''+py)'+py'+qy,\qqq y(0)=y(1)=y(2)=0,
\]
where the 1-periodic coefficients $p,q$ satisfy
\[
\lb{spacegH} \gu=(p,q)\in\gH=\mH\os\mH.
\]
Here  $\mH=L_\R^2(\T),\T=\R/\Z$, is the Hilbert space, equipped with
the norm $\|f\|^2=\int_0^1|f(x)|^2dx$. The spectrum of $\cL$ is pure
discrete \cite{BK21} and it is complex.

The main goal of the present work is to obtain energy-uniform estimates
for perturbation of the spectrum and the norming constants
of the 3-point problem in the case of small coefficients.
These results are necessary for solving the inverse 3-point problem,
which will be the topic of our next work.

McKean \cite{McK81}
considered the Lax operator of the periodic GBE
with the coefficients $p,q\in C^\iy(\T)$. For the case of small coefficients
he obtained some uniqueness results in the inverse
spectral problem. His paper contains many important ideas
and results, but his arguments cannot be directly applied
to the case of non-smooth coefficients.
In our paper \cite{BK21} we considered the operator
$\cL$, given by \er{Hdpq}, with the coefficients $p,q\in L^1(\T)$.
We determined its eigenvalue asymptotics at high energy and
the trace formula for this operator.
Unfortunately, it is not enough to solve the inverse spectral 3-point problem.
We need  sharp uniform asymptotics of eigenvalues and the norming constants
for small coefficients, which are determined here.
Note that we are not aware of any works that
 discuss norming constants
for higher-order operators with multi-point conditions
and where similar uniform estimates were obtained.
Apparently, our results in these directions are completely new.

The Lax operator for the bad periodic Boussinesq equation is
self-adjoint third-order operators with periodic coefficients.
Describe briefly results about it.
The spectrum is absolutely continuous and covers the
real line. It has multiplicity 1 or 3. The spectrum of multiplicity 3
consists of a finite number $\ge 0$ of bounded intervals separated by gaps.
Edges of the gaps are branch points of the Riemann surface.
The branch points are asymptotically located far from the real axis.
A detailed study of the spectral properties of these operators
was begun by the authors  in  \cite{BK12}, \cite{BK14}.
We analyzed the Riemann surface and the spectrum
and determined high energy asymptotics of the branch points and
the 2-periodic spectrum. Moreover, we considered
the case of small coefficients.
In the case of small coefficients either whole spectrum
has multiplicity one or there is exactly one small interval
with the spectrum of multiplicity 3. The asymptotics of this
small interval was determined in \cite{BK12}.
Amour \cite{A99}, \cite{A01} considered self-adjoint third-order operators
on the interval $[0,1]$ under the Dirichlet plus quasi-periodic boundary
conditions. He proved that two spectra uniquely define the operator
and obtained explicit formulas for isospectral flows.

Higher-order operators and operators with matrix periodic
coefficients have numerous applications. For this reason, they have
been the subject of research by mathematicians for many years.
Gelfand \cite{G50} proved the decomposition of periodic operators on
the line into a direct integral. These operators in different
classes of coefficients considered by Badanin and Korotyaev
\cite{BK14x}, \cite{BK15}, Mikhailets and Molyboga \cite{MM04},
Papanicolaou \cite{P95}, \cite{P03}, ...
The general theory of linear systems of
differential equations with periodic coefficients was developed by
Gel'fand and Lidskii \cite{GL55}, Krein \cite{Kr55} and Korotyaev
with co-authors \cite{CK06}, \cite{K10}, see also Carlson
\cite{C00}, Clark, Holden, Gesztesy, and Levitan \cite{CHGL00}.
Korotyaev with coauthors consider the matrix Hill operators
\cite{BBK06}, \cite{CK06}, and the matrix periodic Dirac type
operators \cite{K08}, \cite{K10}. There the Riemann surface
and Lyapunov functions were analyzed, and based on this analysis
the spectral properties of the operator were studied. Moreover,
we constructed a conformal mapping (averaged quasimomentum) and
obtained various properties of this mapping. We defined various
new trace formulas for the potentials and the Lyapunov exponent
and obtained a priori estimates of the gap lengths in terms of the potentials.
Finally, we note that the inverse problem for the Schr\"odinger operator
with matrix-valued potential was solved by Chelkak and Korotyaev
in \cite{CK06q}, \cite{CK09}
via eigenvalues plus matrix-valued norming constants.

There are a huge number of articles on spectral asymptotics
for higher order operators.
Multipoint problems are much less studied,
see the short review in our paper \cite{BK21}. At the same time, apparently,
such problems are important
in the theory of nonlinear completely integrable systems.

Note that the spectral analysis of scalar higher-order operators
with periodic coefficients and the periodic systems
(the first and the second order) has both common properties
and fundamentally different ones,
see \cite{BBK06}, \cite{CK06}, \cite{K08}, \cite{K10}, \cite{BK11}.
Fundamental solutions of the periodic system are uniformly bounded
on the real line, while higher-order equations have exponentially
increasing fundamental solutions. It is necessary to take into account
the contribution of bounded solutions to the spectral asymptotics
against the background of the contribution of increasing solutions.
This requires
a more complicated analysis than the corresponding analysis for the Hill operators.
At the same time, the Riemann surfaces
of the matrix Hill operators at high energy are more complex
than the Riemann surfaces
of higher-order operators with periodic coefficients. This
complicates the spectral analysis of the matrix operators.

The difficulties we encounter in analyzing the operator $\cL$ are
due to two circumstances. First, we are dealing with a higher-order operator.
As has already been said, in order to determine spectral asymptotics
for such operators we have to use more complicated techniques
compared to second-order operators.
Second, $\cL$ is a non-self-adjoint operator. For this reason,
we must simultaneously analyze two differential equations:
the direct one and the transposed one,
which corresponds to the formally adjoint differential expression.

\subsection{Eigenvalues}
Introduce the fundamental solutions $\vp_1, \vp_2, \vp_3$ of the equation
\[
\lb{1b}
(y''+py)'+py'+q y=\l y,\qq\l\in\C,
\]
satisfying the conditions
$\vp_j^{[k-1]}|_{x=0}=\d_{jk}, j,k=1,2,3$,
where $y^{[0]}=y,y^{[1]}=y',y^{[2]}=y''+ py$.
The spectrum $\s(\cL)$ of $\cL$ is pure discrete
and satisfies
\[
\lb{spec}
\s(\cL)=\{\l\in\C:\Delta(\l)=0\},
\]
where the characteristic function $\Delta$ is the entire function given by
\[
\lb{defsi}
\Delta(\l)=\det\ma\vp_2(1,\l)&\vp_3(1,\l)\\
\vp_2(2,\l)&\vp_3(2,\l)\am.
\]
All eigenvalues of $\cL$ at $p=q=0$ are simple, real, and
have the form
\[
\lb{unpev}
\m_{n}^o=(z_n^o)^3,\qq\text{where} \qq z_n^o={2\pi n\/\sqrt3},\qq
n\in\Z_0=\Z\sm\{0\}.
\]

Introduce the Hilbert spaces
\[
\lb{spacegH1}
\mH^1=\mH^1(\T),\qqq
\gH^1=\mH^1\os\mH,
\]
equipped with the norms
\[
\lb{defcH1}
\|\gu\|_{\mH^1}=\|p'\|,\qqq
\|\gu\|_1^2=\|p'\|^2+\|q\|^2,
\]
respectively. Introduce the ball $\cB_1(r)\ss \gH^1$ by
\[
\lb{defcB1ball}
\cB_1(r)=\{\gu\in\gH^1:\|\gu\|_1<r\}, \qqq r>0.
\]
Introduce  the Fourier coefficients
$$
\wh f_n=\int_0^1f(x)e^{-i2\pi nx}dx,\qq
\wh f_{cn}=\int_0^1f(x)\cos2\pi nxdx,\qq
\wh f_{sn}=\int_0^1f(x)\sin2\pi nxdx,\qq n\in\Z.
$$
Introduce the domains $\cD_n$ by
\[
\lb{DomcD}
\cD_{n}=\Big\{\l\in\C:\big|z-z_n^o\big|
<1\Big\},\qq\cD_{-n}=- \cD_{n},\qq n\ge 0,
\]
here and below
$$
z=\l^{1\/3},\qq
\arg z\in\Bigl(-{\pi\/3},{\pi\/3}\Bigr],\qq
\arg\l\in(-\pi,\pi].
$$

Throughout the text below, $\ve>0$ is some fixed, sufficiently small number.
In addition, we will denote by $C$ various positive constants
whose values depend only on $\ve$.
We formulate our first main result.

\begin{theorem}
\lb{Th3pram} Let $\gu\in\cB_1(\ve)$ and let $\wh p_0=\wh q_0=0$.
Then there is exactly one simple real eigenvalue $\m_n$ of the
operator $\cL$ in each domain $\cD_n,n\in\Z_0=\Z\sm\{0\}$, which
satisfies
\[
\lb{asmun}
\Big|\m_n-\m_{n}^o+\ga_n-{\gb_n\/\sqrt3}\Big|\le{C\|\gu\|_1^2\/n},
\]
for some $C>0$, where
\[
\lb{deffn}
\ga_n={\wh p_{sn}'\/\sqrt3}+\wh q_{cn},\qq
\gb_n={\wh p_{cn}'\/\sqrt3}-\wh q_{sn}.
\]

\end{theorem}

\subsection{Norming constants}
The eigenvalues $\m_n,n\in\Z_0$, do not uniquely determine
the coefficients $p$ and $q$.
To recover the coefficients, it is necessary to add
the set of norming constants.

Recall the norming constants for the Schr\"odinger operator from  \cite{PT87}. Consider the operator $-y''+Vy$ on $L^2(0,1)$
under the Dirichlet boundary conditions $y(0)=y(1)=0$, where $V\in
\mH$. Recall that the spectrum consists of the simple eigenvalues
$\gm_1<\gm_2<...$, $\gm_n=(\pi n)^2+O(1)$, as $n\to+\iy$. These
eigenvalues are zeros of an entire function $\vp(1,\cdot)$, where
$\vp(x,E)$ is the fundamental solution of the equation
\[
\lb{hilleq}
-y''+Vy=E y,
\]
satisfying the initial conditions
$\vp|_{x=0}=0,\vp'|_{x=0}=1.$
The norming constant $\gh_{sn}$ associated with the eigenvalue $\gm_n$
is defined by
\[
\lb{ncschr}
\gh_{sn}=2\pi n\log|\vp'(1,\gm_n)|,\qq {\rm where} \ \ \log 1=0.
\]

Introduce the {\it monodromy matrix} $M$ for $\cL$ by
\[
\lb{defmm}
M(\l)=\big(\vp_j^{[k-1]}(1,\l)\big)_{j,k=1}^3.
\]
The matrix-valued function $M$ is entire and real on $\R$.
The characteristic polynomial $D$ of the monodromy matrix is given by
\[
\lb{1c} D(\t,\l)
=\det(M(\l)-\t \1_{3}),\qq (\t,\l)\in\C^2.
\]
An eigenvalue  of $M$ is called a {\it multiplier}, it is a
zero of the polynomial $D(\cdot,\l)$.
The matrix $M$ has exactly $3$ (counting with
multiplicities) eigenvalues $\t_j,j=1,2,3$, which  satisfy
$
\t_1\t_2\t_3=1.
$
In particular, each $\t_j\ne 0$ for all $\l\in\C$.
The multipliers $\t_1,\t_2,\t_3$ constitute three distinct
branches of some function analytic on a 3-sheeted Riemann surface $\cR$.
This surface is an invariant of the Boussinesq flow.
The surface $\cR$ has only algebraic singularities in $\C$.
These singularities are {\it ramifications} of the surface $\cR$.
We consider the ramifications in more detail in our paper \cite{BK24xxxx}.

Define norming constants for the 3rd order operator.
Consider  the transpose operator
\[
\lb{trop}
\wt\cL \wt y=-(\tilde y''+p\tilde y)'-p\tilde y'+q \tilde y,\qqq
\tilde y(0)=\tilde y(1)=\tilde y(2)=0.
\]
Theorem~\ref{Th3pram} yields that if $\gu\in\cB_1(\ve)$,
then there is exactly one simple eigenvalue $\wt\m_n$ of
the  operator  $\wt\cL$ in each domain $\cD_n,n\in\Z_0$.   Let $\wt
y_n(x)$ be a corresponding eigenfunction such that $\wt y_n'(0)=1$.
There exists a multiplier, which  is simple and positive on the half-line $[1,+\iy)$. Denote it by $\t_3$.
Define the norming constants $h_{sn}, n\in \N$, by the identities
\[
\lb{defnf}
 h_{sn} =8(\pi n)^2\log |\wt y_n'(1)\t_3^{-{1\/2}}(\wt\m_n)|, \qq
\t_3^{1\/2}(\wt\m_n)>0.
\]

Consider the McKean transformation, mentioned above, in more detail.
This is a standard transformation that lowers the order of the equation
so that the 3rd order equation \er{1b} becomes
the Hill equation \er{hilleq} with an energy-dependent potential $V=V(E,\gu)$
and $E={3 \/4}\l^{2\/3}$. We consider this transformation in \cite{BK24x}, where  we proved the following results:

{\it Let $(\gu,n)\in\cB_1(\ve)\ts \N$ and let $\gm_n$ be the
eigenvalue of the Dirichlet problem for Eq.~\er{hilleq} with the
potential $V=V(E,\gu)$, and let $\gh_{sn}$ be the corresponding
norming constant,  given by \er{ncschr}. Let $h_{sn}$ be given by
\er{defnf}. Then
\[
\lb{relr3p2o}
\gm_n ={3\/4}\wt\m_n^{2\/3},\qq
\gh_{sn}={h_{sn}\/4\pi n}.
\]
}

In this paper we determine asymptotics of the norming constants,
given by \er{defnf}.

\begin{theorem}
\lb{Thnf}
Let $\gu\in\cB_1(\ve)$ and let $\wh p_0=\wh q_0=0$. Then
\[
\lb{asncr}
\Big|h_{sn}-\ga_n+\sqrt3\gb_n\Big|\le{ C\|\gu\|_1^2\/n},
\]
for all $n\in\N$ and for some $C>0$, where $\ga_n$ and $\gb_n$
are given by \er{deffn}.
\end{theorem}

\no {\bf Remark.}
1) Using the symmetry \er{symev} we can define the norming
constants for $n\le -1$ and extend the estimates \er{asncr} onto this case,
see Corollary~\ref{Cornf}.

\no 2) The sequences $h_{cn}=\m_n-\m_n^o$ and
$h_{sn},n\in\Z_0$, form the set of spectral data of the inverse problem
for the operator $\cL$, see \cite{BK24xx}.

\section{Estimates of the fundamental matrix}
\setcounter{equation}{0}

\subsection{The fundamental matrix}
Introduce the {\it fundamental matrix}
$\Phi(x,\l),(x,\l)\in\R\ts\C$, of Eq.~\er{1b} by
\[
\lb{deM}
\Phi=(\Phi_{jk})_{j,k=1}^3=\ma\vp_1&\vp_2&\vp_3\\
\vp_1'&\vp_2'&\vp_3'\\
\vp_1^{[2]}&\vp_2^{[2]}&\vp_3^{[2]}\am,\qq \Phi|_{x=0} =\1_3,
\]
where $\vp_1, \vp_2, \vp_3$ are the fundamental solutions,
$y^{[2]}=y''+py$,
$\1_3$ is the $3\ts3$ identity matrix.
The fundamental matrix $\Phi$ satisfies the equation
\[
\lb{me1}
\Phi'= H\Phi,\qq \Phi|_{x=0}=\1_3,\qq H=H_0-Q,
\]
where the $3\ts 3$ matrix-valued functions
$H_0, Q$ have the form
\[
\lb{mu}
H_0=\ma 0&1&0\\0 &0&1\\\l&0&0\am,\qq
 Q=\ma 0&0&0\\p&0&0\\q&p&0\am.
\]
The matrix $\Phi$ satisfies the Liouville identity
$$
(\det \Phi)'=\det \Phi\Tr(\Phi'\Phi^{-1})=\det \Phi\Tr H=0,
$$
which yields
$
\det \Phi=1.
$
The standard arguments, see, e.g. \cite[Lm~2.1]{BK14}, show that
each of the matrix-valued function $\Phi(x,\cdot),x\in\R_+$, is entire
and real on $\R$.
Moreover, each of $\Phi(x,\l,\cdot),(x,\l)\in\R_+\ts\C$
is analytic in the Hilbert space $ \gH$  given by \er{spacegH}.

Consider the unperturbed equation $y'''=\l y$.
In this case $Q=0$ and the corresponding fundamental matrix $\Phi_0$
has the form
$\Phi_0=e^{xH_0}. $
The eigenvalues of the matrix $H_0$ are given by
$z,\o z,\o^2 z$, here and below
$$
\o=e^{i{2\pi\/3}}.
$$
Then the eigenvalues of the matrix $\Phi_0$ have the form $e^{zx},e^{\o zx},e^{\o^2 zx}$.

Consider the transpose operator $\wt\cL$, given by \er{trop},
and the corresponding equation
\[
\lb{1btr}
-(\tilde y''+p\tilde y)'-p\tilde y'+q \tilde y=\l \tilde y.
\]
Introduce the fundamental solutions $\tilde\vp_1, \tilde\vp_2, \tilde\vp_3$
to Eq.~\er{1btr} and the fundamental matrix $\tilde \Phi$ by
\[
\lb{deMc}
\tilde \Phi=\ma\tilde\vp_1&\tilde\vp_2&\tilde\vp_3\\
\tilde\vp_1'&\tilde\vp_2'&\tilde\vp_3'\\
\tilde\vp_1^{[2]}&\tilde\vp_2^{[2]}&\tilde\vp_3^{[2]}\am,\qq
\tilde \Phi|_{x=0} =\1_3.
\]

\begin{lemma}
\lb{T21}

 Let $\gu\in \gH$, where $\gH$ is given by \er{spacegH}. Then
the matrix-valued functions $\Phi$ and $\tilde \Phi$ satisfy
\[
\lb{simMt}
\wt \Phi^\top J\Phi=J,\qq \wt \Phi =J(\Phi^\top)^{-1}J,
\qq\text{where}\qq
J=\ma 0&0&1\\0&-1&0\\1&0&0\am.
\]

\end{lemma}

\no {\bf Proof.} Introduce  the matrix coefficient $\tilde H$
for the transpose equation \er{1btr} by
$$
\tilde H=
\ma 0&1&0\\-p &0&1\\q-\l&-p&0\am.
$$
Then $\Phi'=H\Phi$, see \er{me1},
$\tilde \Phi'=\tilde H\tilde \Phi$, and
$\Phi|_{x=0} =\tilde \Phi|_{x=0} =\1_3.
$
Moreover, we have
$J^\top=J$, $J^2=\1_3$, and $J  H=-(J\tilde  H)^\top$.
Using the identities
$$
J\Phi'=J  H \Phi,\qq
(\tilde \Phi')^\top J=\tilde \Phi^\top(J\tilde  H)^\top
=-\tilde \Phi^\top J  H,
$$
we obtain
$$
(\tilde \Phi^\top J\Phi)'=(\tilde \Phi^\top)' J\Phi+\tilde \Phi^\top J\Phi'
=-\tilde \Phi^\top J  H \Phi+\tilde \Phi^\top J  H \Phi=0,
$$
which yields \er{simMt}.~\BBox

\subsection{Transformations of Eq.~\er{me1}}
Our method for obtaining asymptotics uses the usual transition
from differential equations to integral ones. The asymptotic behavior
of the solutions is obtained by iterations in the integral equations.
To improve the convergence of the iterations, we will first
carry out some transformations of the differential equation \er{me1}.

Let $\Psi$ be a solution to Eq.~\er{me1} such that
$\det\Psi\ne 0$.
Then the fundamental matrix satisfies
\[
\lb{MsimPhii}
\Phi(x,\l)=\Psi(x,\l)\Psi^{-1}(0,\l).
\]
Let $\l\in\C\sm\{0\}$. Introduce the matrix $U_1$ by
\[
\lb{defmaZ}
 U_1=\ma1&1&1\\\o z&\o^2 z&z \\
\o^2z^2&\o z^2&z^2\am,\qq\text{then}\qq
 U_1^{-1}=
{1\/3}\ma 1&{\o^2\/z}&{\o\/z^2}\\1&{\o\/z}&{\o^2\/z^2}\\1&{1\/z}&{1\/z^2}\am,
\qq \det U_1=-i3\sqrt3\l.
\]
Introduce the matrix-valued function $Y_1$ by the identity
\[
\lb{defcM}
  Y_1=U_1^{-1} \Psi.
\]

\begin{lemma} Let $(\l,\gu)\in(\C\sm\{0\})\ts\gH$, where $\gH$ is given by
\er{spacegH}.
Then the matrix-valued function $Y_1$, given by \er{defcM},
satisfies the equation
\[
\lb{eqcM}
Y_1'-z\O Y_1={ F_1\/z} Y_1,
\]
where
\[
\lb{4g.Om}
\O=\diag(\o,\o^2,1),
\]
\[
\lb{cK12}
 F_1=-{p\O_1\/3}-{q\O_2\/3z},\qqq
\O_1=\ma \o^2&-\o&-1\\
-\o^2&\o&-1\\
-\o^2&-\o&1\am,
\qq
 \O_2=\ma\o&\o&\o\\\o^2&\o^2&\o^2\\ 1& 1& 1\am.
\]
Moreover,
\[
\lb{detY1'}
(\det Y_1)'=0.
\]
\end{lemma}

\no {\bf Proof.} Substituting \er{defcM}
into Eq.~\er{me1} and using the identity
$$
 U_1^{-1}(H_0- Q) U_1
=z\O+{ F_1\/z},
$$
we obtain Eq.~\er{eqcM},
here $H_0,Q$, and $ U_1$ and are given by \er{defmaZ} and \er{mu}.
The Liouville identity gives
$$
(\det Y_1)'=\det Y_1\Tr\Big(z\O+{ F_1\/z}\Big)=0,
$$
which yields \er{detY1'}.~\BBox

\medskip

Eq.~\er{eqcM} has a diagonal coefficient
on the left side and the coefficient, decreasing as $O({1\/z})$,
on the right side.
If $p',q\in L^2(\T)$, then we can get
the coefficient, decreasing as $O({1\/z^2})$ on the right side of
the differential equation.
Let $\gu\in \gH^1$, where $\gH^1$ is given by \er{spacegH1}.
Introduce
the domain $\L$ in $\C$ by
\[
\lb{defLa}
\L=\{\l\in\C:|\l|>1\}.
\]
Introduce the
matrix-valued function $U_2$ by
\[
\lb{defcU1}
U_2=\1_3+{p\O_3\/3z^2},\qq \det U_2=1,
\]
where
\[
\lb{defW1ss}
\O_3={i\/\sqrt3}\ma 0&\o&-\o\\-\o^2&0&\o^2\\1&-1&0\am.
\]
Introduce the matrix $Y_2$ by the identity
\[
\lb{cMFWss}
Y_2=U_2^{-1}Y_1.
\]
Introduce the matrix-valued function
\[
\lb{defmL1}
 L
=\ma 0&-\o  v&-\o \ol  v\\
-\o^2 \ol  v&0&-\o^2 v\\
- v&-\ol  v&0\am,
\qq
\text{where}
\qq
 v={ip'\/\sqrt3}+q.
\]
In the following lemma we rewrite Eq.~\er{eqcM} in the form \er{me11ss}
 having a diagonal coefficient on the left side and
the coefficient, decreasing as $O({1\/z^2})$ on the right side.
Below we use the standard operator norm of matrices
 $|A|=\sup_{|h|=1}|Ah|,|h|^2=\sum_{j=1}^3|h_j|^2$,

\begin{lemma} Let $(\l,\gu)\in\L\ts\cB_1(\ve)$.
Then the function $Y_2$ satisfies the equation
\[
\lb{me11ss}
Y_2'-z\Theta Y_2={ F_2\/z^2} Y_2,
\]
where
\[
\lb{Phi1Theta}
\Theta=\O-{p\O^{2}\/3z^2}-{q\O\/3z^3}-{pp'\O^{2}\/27z^5},
\qq
 F_2=z^2\Big(U_2^{-1}\big(z\O U_2 +{ F_1 U_2\/z}-U_2'\big)-z\Theta\Big),
\]
$ F_1$ is given by \er{cK12}.
Moreover, the diagonal entries of $ F_2$ vanish:
\[
\lb{Phi1jj=0}
 F_{2,jj}=0,\qq j=1,2,3,
\]
and $ F_2$ satisfies the estimate
\[
\lb{defPwtQbet}
\sup_{x\in\R}\Big| F_2(x,\l)-{ L\/3}\Big|\le
{C\|\gu\|_1^2\/|z|},
\]
for some $C>0$,
where the matrix $ L$ is given by \er{defmL1},
the norm $\|\gu\|_1$ is defined by
\er{defcH1}.
Moreover,
\[
\lb{detY2'}
(\det Y_2)'=0.
\]

\end{lemma}

\no {\bf Proof.}
Substituting the identity \er{cMFWss} into Eq.~\er{eqcM} we obtain
\[
\lb{me3}
Y_2'= A Y_2,
\]
where
\[
\lb{PhicN}
 A=U_2^{-1}(B U_2-U_2'),\qq
B=z\O +{ F_1\/z}=z\O-{p\O_1\/3z}-{q\O_2\/3z^2},
\]
here we used \er{cK12}.
The diagonal entries of the matrix $ A$ has the form
\[
\lb{Phijjpr}
 A_{jj}=z\O_{jj} +{ F_{1,jj}\/z} -(U_2^{-1}U_2')_{jj},\qq j=1,2,3.
\]
The definitions \er{4g.Om} and \er{cK12} imply
\[
\lb{OjjmQjj}
\O_{jj}=\o^j,\qq F_{1,jj}=-{\o^{2j}p\/3}-{\o^jq\/3z},\qq j=1,2,3.
\]
The definition \er{defW1ss} gives
\[
\lb{degW1}
 \O_3^2=-{1\/3}\ma\o^2&\o&1\\\o^2&\o&1\\\o^2&\o&1\am,\qq  \O_3^3=0,
\]
then the definition \er{defcU1} implies
\[
\lb{invU1}
U_2^{-1}=\1_3-{p \O_3\/3z^2}+{p^2 \O_3^2\/9z^4}.
\]
The relations \er{defcU1} and \er{invU1} yield
\[
\lb{U1-1U1'}
U_2^{-1}U_2'
={p' \O_3\/3z^2}-{pp' \O_3^2\/9z^4}.
\]
The relations \er{defW1ss} and \er{degW1} give
\[
\lb{U1-1U1'jj}
(U_2^{-1}U_2')_{jj}
={\o^{2j}pp'\/27z^4},\qq j=1,2,3.
\]
Substituting the identities \er{OjjmQjj} and \er{U1-1U1'jj} into \er{Phijjpr}
we obtain
$$
 A_{jj}=z\o^j -{\o^{2j}p\/3z}-{\o^jq\/3z^2}
-{\o^{2j}pp'\/27z^4}=z\Theta_{jj},\qq j=1,2,3.
$$
The definitions \er{Phi1Theta} and \er{PhicN} give
\[
\lb{idF2}
 F_2=z^2( A-z\Theta).
\]
This yields \er{Phi1jj=0}.
Moreover, Eq.~\er{me3} takes the form \er{me11ss}.

We prove \er{defPwtQbet}.
The definition \er{defcU1} and the identity \er{invU1} give
$$
U_2^{-1}B U_2=\Big(\1_3-{p \O_3\/3z^2}+{p^2 \O_3^2\/9z^4}\Big)B
\Big(\1_3+{p \O_3\/3z^2}\Big)
=B+{p\/3z^2}(B\O_3-\O_3B)+{ \a_1\/z^4},
$$
where $\sup_{x\in\R}|\a_1(x,\l)|\le{C\|\gu\|^2}$,
for some $C>0$. Substituting the definition \er{PhicN}
into this identity we obtain
$$
U_2^{-1}B U_2
=z\O+{A_1\/z}-{q\O_2\/3z^2}+{ \a_2\/z^3},
$$
where  $\sup_{x\in\R}|\a_2(x,\l)|\le{C\|\gu\|^2}$,
for some $C>0$,
\[
\lb{defAj}
A_1={p\/3}(\O\O_3-\O_3\O-\O_1).
\]
Substituting this identity and \er{U1-1U1'} into
the definition \er{PhicN}
we obtain
\[
\lb{Phi1pr}
 A=U_2^{-1}B U_2-U_2^{-1}U_2'
=z\O+{A_1\/z}+{A_2\/z^2}+{ \a_3\/z^3},
\]
where $\sup_{x\in\R}|\a_3(x,\l)|\le{C\|\gu\|_1^2}$,
for some $C>0$,
\[
\lb{defAj1}
A_2=-{1\/3}(q\O_2+p' \O_3).
\]
The definitions \er{4g.Om}, \er{cK12}, \er{defW1ss}, and \er{defmL1} give
$$
\O\O_3-\O_3\O-\O_1+\O^{2}=0,\qq  L=q(\O-\O_2)-p' \O_3.
$$
Substituting these identities
into \er{defAj} and \er{defAj1}  we obtain
$$
A_1=-{p\/3}\O^{2},\qq A_2={ L\/3}-{q\/3}\O.
$$
Then the relations \er{Phi1Theta} and \er{Phi1pr} imply
\[
\lb{idAzO}
 A=z\O-{p\O^{2}\/3z}-{q\O\/3z^2}+{ L\/3z^2}
+{ \a_4\/z^3}
=z\Theta+{ L\/3z^2}+{ \a_5\/z^3},
\]
where $\sup_{x\in\R}|\a_j(x,\l)|\le{C\|\gu\|_1^2},j=4,5$,
for some $C>0$,
$\Theta$ is given by \er{Phi1Theta}.
Then the identities  \er{idF2} and \er{idAzO} yield
$$
 F_2={ L\/3}+{ \a_5\/z}.
$$
This gives \er{defPwtQbet}. The Liouville identity gives
$$
(\det Y_2)'=\det Y_2\Tr\Big(z\Theta+{ F_2\/z^2}\Big)=0,
$$
which yields \er{detY2'}.~\BBox

\subsection{Representation of the fundamental matrix}
Fundamental solutions $\vp_j$ are not very convenient for asymptotic analysis.
Each of them contains both exponentially increasing and exponentially
decreasing at high energies terms. The contribution to the asymptotics
from decreasing terms is lost against the background of increasing ones.
Below, in Lemmas~\ref{LmBest}~ii) and \ref{LmBest1}~ii),
we express the fundamental
matrix in terms of other, more appropriate, fundamental solutions
$\phi_{1,j}$ and $\phi_{2,j}$, given by \er{fsphi}.
The method we will use goes back to Birkhoff, see \cite{N67}.
We formulate our results for $\l\in\C_+$.
In this case we have the simple estimates
\[
\lb{4g.esom}
\Re(\o z)\le\Re(\o^2z)\le\Re z,\qq \forall\ \ \l\in\C_+.
\]
The analytic extension gives
the corresponding results in $\C_-$.

Let $\l\in\L_+$, where $\L_+$ is the domain in $\C_+$ given by
\[
\lb{defLa+}
\L_+=\{\l\in\C_+:|\l|>1\}.
\]
Let $N\in\N$. Consider the Birkhoff integral equation
\[
\lb{ineqX}
 X_1=\1_{3}+{K_1X_1\/z},
\]
on the space $C([0,N])$ of $3\ts 3$ matrix-valued functions
continuous in $x$, where
\[
\lb{defKcX}
(K_1  X_1)_{l j}(x,\l)=\int_0^NK_{1,l j}(x,s,\l)
( F_1   X_1)_{l j}(s,\l)ds,
\qqq l ,j=1,2,3,
\]
\[
\lb{defKie}
K_{1,l j}(x,s,\l)= \ca \ \
e^{z(x-s)(\o^l-\o^j)}\chi(x-s),
\ \ \  l <j\\
-e^{z(x-s)(\o^l-\o^j)}\chi(s-x), \ \ \
l \ge j\ac,\qq \chi(s)=\ca 1,\ s\ge 0\\ 0,\ s<0\ac,
\]
$F_1$ is given by \er{cK12}.
Using the estimates \er{4g.esom}
we obtain
$|K_{1,l j}(x,s,\l)|\le1$ for all
$l,j=1,2,3$, $(x,s,\l)\in[0,N]^2\ts\C_+$.

Introduce the balls $\cB(r),r>0$, in the space $\gH$, having the form \er{spacegH},
by
$$
\cB(r)=\{\gu\in \gH:\|\gu\|<r\},\qq\|\gu\|^2=\|p\|^2+\|q\|^2.
$$
We prove the following results about the integral equation \er{ineqX}.

\begin{lemma}
\lb{LmBest}
Let $N\in\N$. If $(\l,\gu)\in\L_+\ts \cB({1\/6N})$,
then Eq.~\er{ineqX} has a unique solution $ X_1\in C([0,N])$.
Each matrix-valued function $ X_1(\cdot,\l)$,
$\l\in\L_+$,
is continuous on $[0,N]$,
each $ X_1(x,\cdot)$, $x\in[0,N]$,
is analytic on the domain $\L_+$, and satisfies
\[
\lb{bccX0}
 X_{1,lj}(0,\l)= X_{1,jl}(N,\l)=0,\qq 1\le l<j\le 3,\qq
 X_{1,jj}(N,\l)=1,\qq j=1,2,3,
\]
\[
\lb{estcX}
\sup_{x\in[0,N]}| X_1(x,\l)|\le 2,\qq
\sup_{x\in[0,N]}| X_1(x,\l)-\1_3|\le{6N\|\gu\|\/|z|},
\]
for all $\l\in\L_+$.
The function
\[
\lb{defcX}
Y_1= X_1 e^{zx\O}
\]
is the solution to Eq.~\er{eqcM} on the interval $[0,N]$,
satisfying the conditions
\[
\lb{condYij}
Y_{1,lj}(0,\l)=Y_{1,jl}(N,\l)=0,\qq 1\le l<j\le 3,\qq
Y_{1,jj}(N,\l)=e^{N\o^jz},\qq j=1,2,3,
\]
here $\O$ is given by \er{4g.Om}.
The function
\[
\lb{defcA}
\Psi_1=(\Psi_{1,lj})_{l,j=1}^3= U_1 Y_1= U_1 X_1 e^{zx\O}
\]
satisfies Eq.~\er{me1} on the interval $[0,N]$,
here $ U_1$ is given by \er{defmaZ}. Moreover,
\[
\lb{dets1}
\det X_1=\det Y_1=1,\qq\det\Psi_1=-i3\sqrt3\l.
\]
Each of the functions $ X_1^{-1}(\cdot,\l)$,
$\l\in\L_+$, is continuous on $[0,N]$,
each $ X_1^{-1}(x,\cdot)$, $x\in[0,N]$,
is analytic on the domain $\L_+$.
The fundamental matrix satisfies
\[
\lb{factPhi1}
\Phi(x,\l)=\Psi_1(x,\l)\Psi_1^{-1}(0,\l)
=U_1(\l) X_1(x,\l) e^{zx\O}  X_1^{-1}(0,\l)U_1^{-1}(\l),
\]
for all $x\in[0,N]$.
Each of the functions $ X_1(x,\l,\cdot)$, $ X_1^{-1}(x,\l,\cdot)$,
$(x,\l)\in[0,N]\ts\L_+$,
is analytic on the ball $\cB({1\/6N})$.
If, in addition, $\gu\in\cB({1\/12N})$, then
\[
\lb{estcX-1}
\sup_{x\in[0,N]}| X_1^{-1}(x,\l)-\1_3|\le{12N\|\gu\|\/|z|}.
\]

\end{lemma}

\no {\bf Proof.}
Iterations in the integral equation \er{ineqX} give
\[
\lb{secX}
 X_1=\sum_{n=0}^\iy {K_1^n\1_3\/z^n}.
\]
Moreover, the definition \er{defKcX} gives
the estimate
$$
\sup_{x\in[0,N]}|(K_1X)(x,\l)|\le \int_0^N| F_1(s,\l)||X(s,\l)|ds,
$$
for a continuous $3\ts3$ matrix-valued function $X$
and the definitions \er{cK12} imply
$$
\int_0^1| F_1(s,\l)|ds\le3\Big(\|p\|+{\|q\|\/|z|}\Big)
\le3\|\gu\|,
$$
as $|z|\ge1$. Then
$$
\begin{aligned}
\sup_{x\in[0,N]}|K_1^n\1_3(x,\l)|\le\Big(\int_0^N| F_1(s)|ds\Big)^n
\le(3N\|\gu\|)^n,\qqq n\ge 0.
\end{aligned}
$$
Therefore, the series \er{secX} converges absolutely
and uniformly on any bounded subset of
$[0,N]\ts \L_+\ts\cB({1\/3N})$.
Each term is a continuous function of $x\in[0,N]$
and an analytic function of $\l\in\C$ and $\gu\in \gH$,
then the sum is continuous in $x\in[0,N]$ and
analytic in $\l\in\L_+$ and $\gu\in\cB({1\/3N})$.
Summing the majorants in the series \er{secX} we obtain
\[
\lb{estX1X1-1}
\sup_{x\in[0,N]}| X_1(x,\l)|\le{1\/1-A},
\qq
\sup_{x\in[0,N]}| X_1(x,\l)-\1_3|\le {A\/1-A},
\qq A={3N\|\gu\|\/|z|}.
\]
If $\gu\in \cB({1\/6N})$, then
$ A\le{1\/2}$,
which yields \er{estcX}.
The definitions \er{defKcX} and \er{defKie} give
$(K_1 X_1)_{lj}|_{x=0}=0$, $l<j$,
$(K_1 X_1)_{lj}|_{x=N}=0$, $l\ge j$. This gives \er{bccX0}.

In \cite[Cor~5.1]{BK21} we proved that $Y_1$, given by \er{defcX},
is the solution to Eq.~\er{eqcM} on the interval $[0,N]$.
The identities \er{bccX0} provide \er{condYij}.
The identity \er{defcM} implies
\er{defcA}.
The identities \er{detY1'} and \er{condYij}
imply $\det Y_1=\det e^{N\o^jz} =1$.
Then the identities \er{defmaZ} and \er{defcX} give \er{dets1}.
Therefore,
$ X_1^{-1}$ exists and is continuous with respect to $x$ and analytic
with respect to $\l$ and $\gu$.
The relations \er{MsimPhii} and \er{defcA}  give \er{factPhi1}.

The estimate \er{estX1X1-1} shows that
if $\gu\in\cB({1\/12N})$, then
$$
\sup_{x\in[0,N]}| X_1(x,\l)|
\ge1-\sup_{x\in[0,N]}| X_1(x,\l)-\1_3|
\ge1-2A\ge{1\/2},
$$
and
$$
\sup_{x\in[0,N]}| X_1^{-1}(x,\l)-\1_3|
=\sup_{x\in[0,N]}| X_1^{-1}(x,\l)||\1_3- X_1(x,\l)|
\le 2\sup_{x\in[0,N]}|\1_3- X_1(x,\l)|\le
4A.
$$
The definition of $A$ in \er{estX1X1-1} yields \er{estcX-1}.
~\BBox

\subsection{Representation for the case $p',q\in L^1(\T)$}
Let $\gu\in \gH^1$, where $\gH^1$ is given by \er{spacegH1},
let $N\in\N$, and let $K_2$ be an integral operator on
the space $C[0,N]$ of $3\ts 3$ continuous matrix-valued functions
on $[0,N]$ given by
\[
\lb{4g.dcLipr}
(K_2 X_2)_{l j}(x,\l)=\int_0^NK_{2,l j}(x,s,\l)
( F_2  X_2)_{l j}(s,\l)ds
\qqq\forall\qq l ,j=1,2,3,
\]
where
\[
\lb{Kljpr}
K_{2, l j}(x,s,\l)= \ca \ \
e^{z\int_s^x(\Theta_ l(u,\l)-\Theta_j(u,\l))du}\chi(x-s),
\ \ \   l <j\\
-e^{-z\int_x^s(\Theta_ l(u,\l)-\Theta_j(u,\l))du}\chi(s-x), \ \ \
 l \ge j\ac,\qq \chi(s)=\ca 1,\ s\ge 0\\ 0,\ s<0\ac,
\]
$ F_2$ and $\Theta$ are given by \er{Phi1Theta}.
Consider the integral equation
\[
\lb{4g.me5ipr}
 X_2=\1_{3}+{K_2 X_2\/z^2}
\]
on the space $C[0,N]$.
Introduce the matrix
\[
\lb{defL0}
\cN=(\cN_{lj})_{l,j=1}^3,\qqq
\cN_{lj}(x,t,\l)=K_{1,lj}(x,t,\l) L_{lj}(t).
\]
where $K_{1,lj}$ have the form \er{defKie}, $ L$ is given by \er{defmL1}.
We prove the following results.

\begin{lemma}
\lb{LmBest1}
Let $N\in\N$ and let $\ve>0$ be small enough.

i) If $(\l,\gu)\in \L_+\ts\cB_1(\ve)$,
then the integral equation \er{4g.me5ipr}
has a unique solution $ X_2\in C([0,N])$.
Each matrix-valued function $ X_2(\cdot,\l)$,
$\l\in\L_+$,
is continuous on $[0,N]$,
each $ X_2(x,\cdot)$, $x\in[0,N]$,
is analytic on the domain $\L_+$, and
\[
\lb{bccX1}
 X_{2,lj}(0,\l)= X_{2,jl}(N,\l)=0,\qq 0\le l<j\le 3,\qq
 X_{2,jj}(N,\l)=1,\qq j=1,2,3.
\]
Moreover, each matrix-valued function $ X_2(x,\l,\cdot),(x,\l)\in[0,N]\ts\L_+$,
is analytic on $\cB_1(\ve)$.

Let $(\l,\gu)\in \L_+\ts\cB_1(\ve)$.
Then the matrix $ X_2$ satisfies
\[
\lb{estcXimp'}
\sup_{x\in[0,N]}\big| X_2(x,\l)\big|\le C,\qq
\sup_{x\in[0,N]}
\Big| X_2(x,\l)-\sum_{n=0}^{ l}{R_n(x,\l)\/z^{2n}}\Big|
\le {C\|\gu\|_1^{ l+1}\/|z|^{2( l+1)}},\qq  l=0,1,2,...,
\]
for some $C>0$, where
\[
\lb{defmRn}
R_n=K_2^n\1_3.
\]
Each  $ X_2^{-1}(\cdot,\l)$,
$\l\in\L_+$, is continuous on $[0,N]$,
each $ X_2^{-1}(x,\cdot)$, $x\in[0,N]$,
is analytic on the domain $\L_+$.
Furthermore,
\[
\lb{estKI3}
\sup_{x\in[0,N]}\Big|R_1(x,\l)
-{1\/3}\int_0^N\cN(x,s,\l)ds\Big|
\le{C\|\gu\|_1^2\/|z|},
\]
for some $C>0$,
where $\cN$ is given by \er{defL0}.

ii) Let $(\l,\gu)\in \L_+\ts\cB_1(\ve)$.
 Then the function
\[
\lb{defY1lm}
Y_2(x,\l)= X_2(x,\l) e^{z\int_0^x\Theta(s,\l)ds}
\]
is the solution to Eq.~\er{me11ss} on the interval $[0,N]$,
satisfying the conditions
\[
\lb{bccY1}
Y_{2,lj}(0,\l)=Y_{2,jl}(N,\l)
=0,\qq 1\le l<j\le 3,\qq
Y_{2,jj}(N,\l)=e^{z\int_0^N\Theta_j(s,\l)ds},\qq j=1,2,3.
\]
The matrix-valued function
\[
\lb{idPhiss}
\Psi_2(x,\l)=\big(\Psi_{2,lj}(x,\l)\big)_{l,j=1}^3
= U_1(\l) U_2(x,\l) X_2(x,\l)
e^{z\int_0^x\Theta(s,\l)ds},\qq x\in [0,N],
\]
satisfies Eq.~\er{me1},
here  $ U_1$  is given by \er{defmaZ}, $U_2$ has the form \er{defcU1}.
 Moreover,
\[
\lb{dets2}
\det X_2=\det Y_2=1,\qq\det\Psi_2=-i3\sqrt3\l.
\]
The fundamental matrix satisfies
\[
\lb{factPhi2}
\Phi(x,\l)=\Psi_2(x,\l)\Psi_2^{-1}(0,\l)
=U_1(\l) U_2(x,\l) X_2(x,\l)
e^{z\int_0^x\Theta(s,\l)ds}  X_2^{-1}(0,\l) U_2^{-1}(0,\l)U_1^{-1}(\l),
\]
for all $x\in[0,N]$.
\end{lemma}

\no {\bf Proof.}
i) Let $(\l,\gu)\in \L_+\ts\cB_1(\ve)$ for some $\ve>0$ small enough.
The definition of $\Theta$ in \er{Phi1Theta}
yields that the kernel $K_{2,lj}(x,s,\l)$ of the integral operator $K_2$,
given by \er{4g.dcLipr}, is bounded, then
$$
\sup_{x\in[0,N]}|K_2 X_2(x,\l)|\le C\|\gu\|_1\int_0^N| X_2(x,\l)|dx,
$$
for some $C>0$.
Repeating the arguments from the proof of Lemma~\ref{LmBest}
we obtain that Eq.~\er{4g.me5ipr}
has a unique solution $ X_2\in C([0,N])$. The definitions
\er{4g.dcLipr} and \er{Kljpr} yield $(K_2 X_2)_{lj}|_{x=0}=0,l<j$,
$(K_2 X_2)_{lj}|_{x=N}=0,l\ge j$. This gives \er{bccX1}.
Iterations in Eq.~\er{4g.me5ipr} give
$$
 X_2=\sum_{n=0}^\iy {R_n\/z^{2n}},
$$
the series converges absolutely and uniformly on any bounded subset of
$[0,N]\ts\L_+\ts\cB_1(\ve)$ and the estimates \er{estcXimp'} hold true.
Each term is continuous in $x\in[0,N]$ and analytic in $\l\in\L_+$
and $\gu\in\cB_1(\ve)$, then the sum is also continuous in $x$
and analytic in $\l$ and $\gu$.

Let $l,j=1,2,3$. Then the definitions \er{4g.dcLipr},
\er{defmRn}, and the estimate \er{defPwtQbet} imply
\[
\lb{K1-K1L}
R_{1,lj}(x,\l)
={1\/3}\int_0^NK_{2,lj}(x,s,\l) L_{lj}(s)ds+{\a_{1}(x,\l)\/z},\qq
\sup_{x\in[0,N]}\big|\a_{1}(x,\l)\big|\le C\|\gu\|_1^2,
\]
for some $C>0$, where $ L$ is given by \er{defmL1}.
The definition \er{Phi1Theta} implies
$$
\Theta_l-\Theta_j
=\o^l-\o^j+{\a_2\/z^2},\qq
\sup_{x\in[0,N]}|\a_2(x,\l)|\le C\|\gu\|.
$$
Then the definitions \er{defKie} and \er{Kljpr} yield
$$
K_{2,l j}=K_{1,l j}+{\a_3\/z},
$$
where $\sup_{(x,s)\in[0,N]^2}|\a_3(x,s,\l)|\le C\|\gu\|$.
Substituting this asymptotics into \er{K1-K1L} and using \er{defL0}
we obtain \er{estKI3}.

ii) In \cite[Cor~6.2]{BK21} we proved that $Y_2$, given by \er{defY1lm},
satisfy Eq.~\er{me11ss} on the interval $[0,N]$.
The identities \er{bccX1} provide \er{bccY1}.
The definitions \er{defcM} and \er{cMFWss} imply
\er{idPhiss}.
The identities \er{detY2'} and \er{bccY1}
yields $\det Y_2=\det e^{z\int_0^N\Theta_j(s,\l)ds} =1$.
Then the identities \er{defmaZ}, \er{defcU1}, and \er{defY1lm}  give \er{dets2}.
The relations \er{MsimPhii} and \er{idPhiss} imply \er{factPhi2}.
~\BBox

\section{3-point eigenvalues}
\setcounter{equation}{0}

\subsection{Two representations of the characteristic function $\D$}
Assume $N=2$ in Lemmas~\ref{LmBest}, \ref{LmBest1}.
Introduce the fundamental solutions
$\phi_{\n,j}(x,\l)$, $\n=1,2$, $j=1,2,3$, $(x,\l)\in[0,2]\ts\L_+$,
to Eq.~\er{1b} by
\[
\lb{fsphi}
\phi_{\n,j}= \Psi_{\n,1j},
\]
where the matrix-valued function $\Psi_1$ is given by \er{defcA},
 $\Psi_2$ is given by \er{idPhiss}.
 Each function $ X_\n(x,\cdot),\n=1,2,x\in[0,2]$,
 from Lemmas~\ref{LmBest}, \ref{LmBest1}, is analytic on $\L_+$,
then $\phi_{\n,j}(x,\cdot),j=1,2,3$, are also analytic.

Introduce the matrix-valued functions $\phi_\n,\n=1,2$, by
\[
\lb{defphiz}
\phi_\n(\l)=\ma
\phi_{\n,1}(0,\l)&\phi_{\n,2}(0,\l)&\phi_{\n,3}(0,\l)\\
\phi_{\n,1}(1,\l)&\phi_{\n,2}(1,\l)&\phi_{\n,3}(1,\l)\\
\phi_{\n,1}(2,\l)&\phi_{\n,2}(2,\l)&\phi_{\n,3}(2,\l)
\am,\qq \l\in\L_+,
\]
where $\phi_{\n,k}$ are given by \er{fsphi}.

\begin{lemma} Assume that $\gu\in\cB(\ve)$ and $\n=1$ or
$\gu\in\cB_1(\ve)$ and $\n=2$. Then
the function $\det\phi_\n$ is entire and satisfies
\[
\lb{idsigma}
\det\phi_\n=-i3\sqrt3\l\Delta,
\]
where the entire function $\Delta$ is given by \er{defsi}.
\end{lemma}

\no {\bf Proof.}
We have the following simple identity
$$
\det\phi_\n(\l)=\det\Psi_\n(0,\l)\Delta(\l),\qq \n=1,2,
$$
see, e.g., \cite[Lm~3.2]{BK21}. Then the identities \er{dets1} and \er{dets2}
give \er{idsigma}.~\BBox

\medskip

In the unperturbed case $p=q=0$ we have $\phi_1=\phi_2=\phi_0$, where
\[
\lb{defphi0}
\phi_0=\ma
1&1&1\\e^{\o z}&e^{\o^2z}&e^{z}\\e^{2\o z}&e^{2\o^2z}&e^{2z}
\am.
\]
The function $\D$ has the form
\[
\lb{ascA0}
\Delta_0
={8\/3\sqrt3\l}\sin{\sqrt3z\/2}\sin{\sqrt3\o z\/2}\sin{\sqrt3\o^2z\/2}
={4\/3\sqrt3\l}\sin\Big({\sqrt3z\/2}\Big)
\lt(\cosh\Big({3z\/2}\Big)
-\cos\Big({\sqrt3z\/2}\Big)\rt).
\]

\subsection{Asymptotics of the fundamental solutions $\phi_{\n}$}

Below we will use the following simple identities
\[
\lb{o-o}
\begin{aligned}
\o+2\o^2=-\o-2=\o^2-1=i\sqrt3\o,\qq
\o^2+2=-\o^2-2\o=1-\o=i\sqrt3\o^2,
\\
1+2\o=-1-2\o^2=\o-\o^2=i\sqrt3.
\end{aligned}
\]
Introduce the functions
\[
\lb{wtbetaj1}
\zeta_j(x,\l)=\sum_{l=1}^3 \int_0^2\cN_{lj}(x,t,\l)dt,\qq j=1,2,3,
\]
where the matrix $\cN$ is given by \er{defL0}.
Recall that
$|K_{1, l j}(x,t,\l)|\le1$, then the definition \er{defL0} gives
\[
\lb{zetaogr}
\sup_{(x,\l)\in[0,2]\ts\C_+}|\zeta_j(x,\l)|\le C,
\]
for some $C>0$ and for all $j=1,2,3$.
Substituting the definitions \er{defmL1}
and \er{defKie} into the definition \er{defL0}  and using the identities \er{o-o}
we obtain
\[
\lb{idL0}
\cN=
\ma 0&-\o e^{i\sqrt3z(x-t)}\chi(x-t) v(t)
&-\o e^{-i\sqrt3\o^2z(x-t)}\chi(x-t)\ol  v(t)\\
\o^2 e^{-i\sqrt3z(x-t)}\chi(t-x)\ol  v(t)
&0&-\o^2e^{i\sqrt3\o z(x-t)}\chi(x-t) v(t)\\
e^{i\sqrt3\o^2z(x-t)}\chi(t-x) v(t)
&e^{-i\sqrt3\o z(x-t)}\chi(t-x)\ol  v(t)&0\am,
\]
where $ v$ is given by \er{defmL1}.
The identities \er{idL0} and the definitions \er{wtbetaj1} give
\[
\lb{wtbetaj}
\begin{aligned}
\zeta_1(x,\l)=\int_x^2e^{i\sqrt3\o^2z(x-s)}{ v}(s)ds
+\o^2\int_x^2e^{-i\sqrt3z(x-s)}\ol  v(s)ds,
\\
\zeta_2(x,\l)=-\o \int_0^xe^{i\sqrt3z(x-s)}{ v}(s)ds
+\int_x^2e^{-i\sqrt3\o z(x-s)}\ol  v(s)ds,
\\
\zeta_3(x,\l)=
-\o^2\int_0^xe^{i\sqrt3\o z(x-s)}{ v}(s)ds
-\o \int_0^xe^{-i\sqrt3\o^2z(x-s)}\ol  v(s)ds.
\end{aligned}
\]
We determine asymptotics of $\phi_{\n,j}$.

\begin{lemma}
i) Let $(\l,\gu)\in\L_+\ts\cB({1\/12})$. Then
the fundamental solutions $\phi_{1,j}$ satisfy
\[
\lb{purtphi}
\sup_{x\in[0,2]}|e^{-zx\o^j}\phi_{1,j}(x,\l)-1|\le{36\|\gu\|\/|z|},
\qq j=1,2,3.
\]

ii) Let $(\l,\gu)\in \L_+\ts\cB_1(\ve)$ and let $j=1,2,3$.  Then
the solutions $\phi_{2,j}$ satisfy
\[
\lb{idphijz}
\phi_{2,j}(x,\l)=\Big(1+{\o^jp(x)\/3z^2}+{w_j(x,\l)\/3z^2}\Big)
e^{z\int_0^x\Theta_{j}(s,\l)ds},\qq x\in [0,2],
\]
where
\[
\lb{defwj}
\sup_{x\in[0,2]}|w_j(x,\l)-\zeta_j(x,\l)|\le{ C\|\gu\|_1^2\/|z|},
\]
for some $C>0$.

\end{lemma}

\no {\bf Proof.}
i) Let $(\l,\gu)\in\L_+\ts\cB({1\/3})$.
The definition \er{defmaZ} gives $ U_{1,1j}=1, j=1,2,3$.
The definitions \er{defcA} and \er{fsphi}
give
\[
\lb{idphj}
\phi_{1,j}=e^{zx\o^j}\sum_{k=1}^3 U_{1,1k} X_{1,kj}
= e^{zx\o^j}\sum_{k=1}^3 X_{1,kj},\qq j=1,2,3.
\]
The relations \er{estcX} and \er{idphj} give
\[
\lb{relcX1}
\phi_{1,j}= e^{zx\o^j}\sum_{k=1}^3\big(\d_{kj}+\alpha_{kj}\big)
= e^{zx\o^j}\big(1+\alpha_j\big),\ \
|\alpha_{kj}(x,\l)|\le{12\|\gu\|\/|z|},\ \ \ \alpha_j=\sum_{k=1}^3\alpha_{kj},
\ \ \ j,k=1,2,3.
\]
The relations \er{relcX1} give \er{purtphi}.

ii) Let $(\l,\gu)\in \L_+\ts\cB_1(\ve)$.
Then the identity \er{idPhiss} gives
\[
\lb{idphijz1}
\phi_{2,j}(x,\l)=\sum_{l=1}^3 \big(U_2(x,\l) X_2(x,\l)\big)_{lj}
e^{z\int_0^x\Theta_{j}(s,\l)ds},\qq x\in [0,2],
\]
where $U_2$ is given by \er{defcU1}.
The definition \er{defcU1} and the estimate \er{estcXimp'}
imply
$$
\sup_{x\in[0,2]}\Big|U_2(x,\l) X_2(x,\l)
-\1_3-{1\/z^2}\Big({p(x)\/3} \O_3+R_1(x,\l)\Big)\Big|
\le{C\|\gu\|_1^2\/|z|^4},
$$
for some $C>0$.
The estimates \er{estKI3} give
\[
\lb{estajpr}
\sup_{x\in[0,2]}\Big|\sum_{l=1}^3 \big(U_2(x,\l) X_2(x,\l)\big)_{lj}
-1-{1\/3z^2}\Big(p(x)\sum_{l=1}^3  \O_{3,lj}
+\sum_{l=1}^3 \int_0^2\cN_{lj}(x,s,\l)ds\Big)\Big|
\le{C\|\gu\|_1^2\/|z|^3},
\]
for some $C>0$, where the matrix $\cN$
is given by \er{defL0}.
The definition \er{defW1ss} and the identities \er{o-o} imply
$
\sum_{l=1}^3  \O_{3,lj}=\o^j, j=1,2,3.
$
The definitions \er{wtbetaj1} and the relations
\er{idphijz1}, \er{estajpr} give
\er{idphijz} and \er{defwj}.~\BBox

\subsection{Counting results}
Introduce the domain
\[
\lb{dmD}
\mD=\L_+\sm\cup_{n\in\N}\ol{\cD_{-n}}.
\]
Recall the standard estimate from \cite[Lm~2.1]{PT87}:
\[
\lb{sePT}
|\sin z|>{1\/4}e^{|\Im z|},\qq\text{as}\qq |z-\pi n|\ge{\pi\/4}\ \ \forall\
n\in\Z.
\]
We prove the following results.

\begin{lemma}
\lb{Lm3-pev}
i) The following estimates hold true
\[
\lb{uest01-3}
|e^{(\o-1)z}|\le e^{-{3\/2} },\qq \forall\ \ \l\in\L_+,
\]
\[
\lb{lesto2-3}
|e^{(\o^2-1)z}-1|>{1\/3},\qq \forall\ \ \l\in\mD.
\]

ii) The function $\Delta_0$, given by \er{ascA0}, satisfies
\[
\lb{estdetphi0}
|\Delta_0(\l)|>
{|e^{z(\o^2+2)}|\/24\sqrt3|\l|},\qq
\l\in\C_+\sm\cup_{n\in\Z}\cD_{n}.
\]

iii) Let $\gu\in\cB(\ve)$.
Then
\[
\lb{estwtcC}
\bigg|{\Delta(\l)\/\Delta_0(\l)}-1\bigg|\le {C\|\gu\|\/|z|},
\]
for all $\l\in\C_+\sm\cup_{n\in\Z}\cD_{n}$ and for some $C>0$.

\end{lemma}

\no {\bf Proof.}
i) If $\l\in\L_+$,
then $|z|\ge 1$ and \er{o-o} implies
$$
|e^{(\o-1)z}|=|e^{-i\sqrt3\o^2z}|=e^{-\Re i\sqrt3\o^2z}=e^{-{3x+\sqrt3y\/2}}
= e^{-\sqrt3 |z|\sin(\vp+{\pi\/3})}
\le e^{-\sqrt3 \sin(\vp+{\pi\/3})}
\le e^{-{3\/2} },
$$
where $z=x+iy=|z|e^{i\vp},0\le\vp\le{\pi\/3}$. This yields \er{uest01-3}.

The definitions \er{4g.Om} and the identities \er{o-o} imply
$\o^2-1=i\sqrt3\o=e^{-i{2\pi\/3}}-1=-\sqrt3e^{i{\pi\/6}}$, then
$$
|e^{(\o^2-1)z}-1|=|e^{i\sqrt3\o z}-1|
=2|e^{i{\sqrt3\/2}\o z}|\Big|\sh\Big( i{\sqrt3\/2}\o z\Big)\Big|
=2|e^{-{\sqrt3\/2}e^{i{\pi\/6}}z}|
\Big|\sh\Big( {\sqrt3\/2}e^{i{\pi\/6}}z\Big) \Big|.
$$
Using the estimate
$$
|e^{-{\sqrt3\/2}e^{i{\pi\/6}}z}|=
e^{-{\sqrt3\/2}\Re(e^{i{\pi\/6}}z)}
=e^{-{\sqrt3\/4}(x\sqrt3-y)}\ge e^{-{3\/4}x}\ge
e^{-{3\/8}}>{2\/3},
$$
we obtain
$$
|e^{(\o^2-1)z}-1|\ge{4\/3}\Big|\sh\Big({\sqrt3\/2}e^{i{\pi\/6}}z\Big)\Big|.
$$
Let, in addition, $\l\in\mD$.
Then the estimate \er{sePT} implies
$|\sh({\sqrt3\/2}e^{i{\pi\/6}}z)|>{1\/4}$, which yields \er{lesto2-3}.

ii) The estimate \er{sePT} gives
$$
\Big|\sin{\sqrt 3 z\/2}\Big|>{1\/4}e^{{\sqrt 3\/2}|\Im z|},\qq
\Big|\sin{\sqrt 3 \o z\/2}\Big|>{1\/4}e^{{\sqrt 3\/2}|\Im \o z|},\qq
\Big|\sin{\sqrt 3 \o^2 z\/2}\Big|>{1\/4}e^{{\sqrt 3\/2}|\Im \o^2 z|},
$$
as $\l\in\C\sm\cup_{n\ge 0}\cD_{n}$.
The identity \er{ascA0} gives
$$
|\Delta_0(\l)|>{e^{{\sqrt 3\/2}(|\Im z|+|\Im \o z|+|\Im \o^2 z|)}\/24\sqrt3|\l|}
={e^{{\sqrt 3\/2}(\sqrt3\Re z+\Im z)}\/24\sqrt3|\l|}=
{|e^{z(\o^2+2)}|\/24\sqrt3|\l|},
\qq \l\in\C_+\sm\cup_{n\in\Z}\cD_{n},
$$
which yields \er{estdetphi0}.

iii) Let $(\l,\gu)\in\L_+\ts\cB(\ve)$.
Substituting the estimates \er{purtphi} into the definition \er{defphiz}
we obtain
\[
\lb{idphi0}
\phi_1=\ma 1+\a_{11}
&1+\a_{12}
&1+\a_{13}\\
e^{\o z}(1+\a_{21})
&e^{\o^2z}(1+\a_{22})
&e^{z}(1+\a_{23})\\
e^{2\o z}(1+\a_{31})
&e^{2\o^2 z}(1+\a_{32})
&e^{2z}(1+\a_{33})\am,\qq
|\a_{jk}(\l)|\le{C\|\gu\|\/|z|},\qq j,k=1,2,3,
\]
for some $C>0$.
The identity \er{idsigma} yields
$$
\Delta={i\det\phi_j\/3\sqrt3\l}
={ie^{(\o^2+2)z}\/3\sqrt3\l}\det\ma 1+\a_{11}&e^{-\o^2z}(1+\a_{12})&e^{-2z}(1+\a_{13})\\
e^{\o z}(1+\a_{21})&1+\a_{22}&e^{-z}(1+\a_{23})\\
e^{2\o z}(1+\a_{31})&e^{\o^2 z}(1+\a_{32})&1+\a_{33}\am.
$$
The estimates \er{4g.esom} yield
$$
\max\{\Re(\o-\o^2)z,\Re(\o^2-1)z,\Re(2\o-2)z,
\Re(\o+\o^2-2)z,\Re(2\o-\o^2-1)z\}\le 0,
$$
then the estimates \er{idphi0} give
\[
\lb{pertmphi}
\Delta={ie^{(\o^2+2)z}\/3\sqrt3\l}\left(\det\ma 1&e^{-\o^2z}&e^{-2z}\\
e^{\o z}&1&e^{-z}\\
e^{2\o z}&e^{\o^2 z}&1\am+A\right)
=\Delta_0+{ie^{(\o^2+2)z}A\/3\sqrt3\l},\qq
|A(\l)|\le{C\|\gu\|\/|z|},
\]
for some $C>0$.
This identity  yields
$$
{\Delta\/\Delta_0}-1
={ie^{(\o^2+2)z}A\/3\sqrt3\l\Delta_0}.
$$
Then the estimates
\er{estdetphi0} and \er{pertmphi} give
 \er{estwtcC}.~\BBox

\medskip
Now we prove counting results for the three-point eigenvalues.

\begin{lemma}
\lb{Th3prami}
Let $\gu\in\cB(\ve)$.
Then there is exactly one simple real
eigenvalue $\m_n$ of the operator $\cL$, given by \er{Hdpq},
and exactly one simple real
eigenvalue $\wt\m_n$ of the operator $\wt\cL$, given by \er{trop},
in each domain $\cD_n,n\in\Z_0=\Z\sm\{0\}$. There are no other eigenvalues.
Moreover,
\[
\lb{symev}
\wt\m_n(\gu)=-\m_{-n}(\gu_*)=\m_{n}(\gu^-),\qq\forall\ \ n\in\Z_0,
\]
where
\[
\lb{epsstar}
\gu_*=(p,-q),\qq
\gu^-(x)=\gu(1-x),\qq
\gu_*^-(x)=\gu_*(1-x),\ \ x\in\R.
\]

\end{lemma}

\no {\bf Proof. }
Let $\l$ belong to the contour
$\pa\cD_n$ for some $n\in\Z$ and let
$\gu\in\cB(\ve)$.
The estimate \er{estwtcC}
and the symmetry $\Delta(\ol\l)=\ol{\Delta(\l)}$ imply
$$
\bigg|{\Delta(\l)\/\Delta_0(\l)}-1\bigg|<1.
$$
This yields
$$
|\Delta(\l)-\Delta_0(\l)|=|\Delta_0(\l)|\bigg|{\Delta(\l)\/\Delta_0(\l)}-1\bigg|
<|\Delta_0(\l)|
$$
on all contours. Hence, by Rouch\'e's theorem,
$\Delta$ has as many zeros, as $\Delta_0$ in each of the
domain  $\cD_n,n\in\Z$.
Since $\Delta_0$ has exactly one simple zero
in each $\cD_n,n\in\Z\sm\{0\}$, the function
$\Delta$ has exactly one simple zero $\mu_n$
in each domain $\cD_n,n\in\Z\sm\{0\}$ and there are no the zero in $\cD_0$.
The zeros are simple and the function $\Delta$ is real on $\R$,
then all zeros are real.
Combining this result with the result of \cite[Lm~5.2~ii]{BK21}, we obtain
that there are no other eigenvalues.

The transpose operator, given by \er{trop},
is equal
to the operator $-\cL(\gu_*)$, then the spectra satisfy
$\s(\wt\cL(\gu))=-\s(\cL(\gu_*))$. Using $\wt\m_n(\gu)\in\cD_n$
and $-\m_{-n}(\gu_*)\in\cD_n$ for all $n\in\Z_0$, we obtain the first
identity in \er{symev}.
Moreover, the operator $\wt\cL(\gu^-)$ is unitarily equivalent
to the operator $\cL(\gu)$, then
$\s(\wt\cL(\gu^-))=\s(\cL(\gu))$. Using $\wt\m_n(\gu^-)\in\cD_n$
and $\m_{n}(\gu)\in\cD_n$ for all $n\in\Z_0$, we obtain
$\wt\m_n(\gu^-)=\m_{n}(\gu)$, which yields
the second identity in \er{symev}.~\BBox

\subsection{Improved asymptotics of the function $\Delta$}
Introduce the matrix-valued function $w(\l),\l\in\L_+$, by
\[
\lb{defwl}
w(\l)=(w_{jk}(\l))_{j,k=1}^3=
\ma
w_1(0,\l)&w_2(0,\l)&w_3(0,\l)\\
w_1(1,\l)&w_2(1,\l)&w_3(1,\l)\\
w_1(2,\l)&w_2(2,\l)&w_3(2,\l)
\am,
\]
where $w_j$ are given by \er{idphijz}.
Define the function
\[
\lb{defWW}
W=\det\ma w_{11}&1&1\\
w_{21}e^{\o z}&e^{\o^2z}&e^{z}\\
w_{31}e^{2\o z}&e^{2\o^2z}&e^{2z}\am
+\det\ma 1&w_{12}&1\\e^{\o z}&w_{22}e^{\o^2z}&e^{z}\\
e^{2\o z}&w_{32}e^{2\o^2z}&e^{2z}\am
+\det\ma1&1&w_{13}\\e^{\o z}&e^{\o^2z}&w_{23}e^{z}\\
e^{2\o z}&e^{2\o^2z}&w_{33}e^{2z}\am.
\]
The functions $w$ and $W$ are analytic on $\L_+$.

\begin{lemma}
Let $(\l,\gu)\in\L_+\ts\cB_1(\ve)$ and let  $\wh p_0=\wh q_0=0$.
Then the function $\Delta$ satisfies
\[
\lb{iddetphi234}
\Delta=\Delta_0+{iW\/9\sqrt3z^5},
\]
where $\Delta_0$ is given by \er{defphi0}.
\end{lemma}

\no {\bf Proof.}
Let $(\l,\gu)\in\L_+\ts\cB_1(\ve)$ and let  $\wh p_0=\wh q_0=0$.
Then $\int_0^1\Theta(s,\l)ds=\O$ and the identities \er{idphijz} yield
\[
\lb{phi2id}
\phi_2=\phi_0+{1\/3z^2}
\left(p(0)\phi_0\O+\ma
w_{11}
&w_{12}
&w_{13}\\
w_{21}e^{\o z}&
w_{22}e^{\o^2z}&
w_{23}e^{z}\\
w_{31}e^{2\o z}&
w_{32}e^{2\o^2z}&
w_{33}e^{2z}\am\right),
\]
where $\O$ is given by \er{4g.Om}, $\phi_0$ has the form \er{defphi0}.
The identities \er{phi2id} and direct calculations give
$$
\det\phi_2=\det\phi_0+{1\/3z^2}\big(p(0)\det\phi_0\Tr\O
+W\big).
$$
The identities $\Tr\O=0$ and \er{idsigma} imply \er{iddetphi234}.~\BBox

Now we improve the asymptotics \er{estwtcC} of the function $\Delta$.
The asymptotics will be expressed in terms of the function $\b$
that we will now define.
Introduce the matrix-valued function $\zeta(\l),\l\in\C$, by
$$
\zeta(\l)=(\zeta_{jk}(\l))_{j,k=1}^3=
\ma
\zeta_1(0,\l)&\zeta_2(0,\l)&0\\
\zeta_1(1,\l)&\zeta_2(1,\l)&\zeta_3(1,\l)\\
0&\zeta_2(2,\l)&\zeta_3(2,\l)
\am,
$$
where $\zeta_j$ are given by \er{wtbetaj1}
(note that, due to \er{wtbetaj}, $\zeta_3(0,\l)=\zeta_1(2,\l)=0$).
Define the function
\[
\lb{defbeta}
\b=
\det\ma
\zeta_{11}&1&1\\
\zeta_{21}e^{\o z}&
e^{\o^2z}&
e^{z}\\
0&
e^{2\o^2z}&
e^{2z}\am+\det\ma
1&\zeta_{12}&1\\
e^{\o z}&
\zeta_{22}e^{\o^2z}&
e^{z}\\
e^{2\o z}&
\zeta_{32}e^{2\o^2z}&
e^{2z}\am+\det\ma
1&1&0\\
e^{\o z}&
e^{\o^2z}&
\zeta_{23}e^{z}\\
e^{2\o z}&
e^{2\o^2z}&
\zeta_{33}e^{2z}\am.
\]
The functions $\zeta$ and $\b$ are analytic on $\C\sm(-\iy,0]$.

Introduce the functions $\theta_j(\l),j=1,...,6$, by
\[
\lb{a145}
\theta_j=\int_0^1e^{-i\sqrt3\o^j zs}{ v}(s)ds,\ \ j=1,2,3,\qq
\theta_4(\l)
=\ol{\theta_1(\ol\l)},\ \
\theta_5(\l)
=\ol{\theta_2(\ol\l)},\ \
\theta_6(\l)
=\ol{\theta_3(\ol\l)}.
\]
The functions $\theta_j(\l),j=1,...,6$, are analytic on $\C\sm(-\iy,0]$.
The definitions \er{wtbetaj} and the identities
$\theta_4=\ol \theta_1,\theta_5=\ol \theta_2,\theta_6=\ol \theta_3$ yield
on $\R_+$:
{\small
\[
\lb{idzeta}
\zeta=\ma
(1+e^{-i\sqrt3\o^2z})\theta_2+\o^2(1+e^{i\sqrt3z})\ol \theta_3&
(1+e^{i\sqrt3\o z})\ol \theta_2&0\\
\theta_2+\o^2\ol \theta_3&\ol \theta_2-\o e^{i\sqrt3z}\theta_3&
-2\Re(\o^2e^{i\sqrt3\o z}\theta_1)\\
0&-\o e^{i\sqrt3z}(1+e^{i\sqrt3z})\theta_3&
-2\Re\big(\o^2e^{i\sqrt3\o z}(1+e^{i\sqrt3\o z})\theta_1\big)
\am\!.
\]
}

The identities
$i\sqrt3\o z_n^o=-(\sqrt3+i)\pi n$ and $i\sqrt3\o^2z_n^o=(\sqrt3-i)\pi n$ give
\[
\lb{thetamn0}
\theta_1(\m_{n}^o)=v_n^+,\qq
\theta_2(\m_{n}^o)= v_n^-,\qq
\theta_3(\m_{n}^o)=\wh  v_n,
\]
where
\[
\lb{whgvin+}
\wh  v_n=\int_0^1e^{-i2\pi ns} v(s)ds,\qq
 v_n^\pm=\int_0^1e^{(\pm\sqrt3+i)\pi ns}v(s)ds.
\]
We prove the following results.

\begin{lemma}
i) The function $\b$, given by \er{defbeta}, satisfies
\[
\lb{idbeta}
\begin{aligned}
\b=
2i\Im\Big(\big(1-e^{-i\sqrt3\o z}+e^{i\sqrt3\o^2z}-e^{-i\sqrt3z}\big)\theta_2
+\o \big(1-e^{-i\sqrt3\o z}-e^{-i\sqrt3\o^2 z}+e^{i\sqrt3z}\big)\theta_3
\\
-2e^{i\sqrt3z}\Re(\o^2e^{i\sqrt3\o z}\theta_1)
-2e^{i\sqrt3\o^2z}\Re\big(\o^2e^{i\sqrt3\o z}(1+e^{i\sqrt3\o z})\theta_1\big)\Big),
\end{aligned}
\]
on $\R_+$,
where $\theta_j$ are given by \er{a145}.
Moreover,
\[
\lb{betaatmn0}
\b(\m_{n}^o)
=i2(-1)^{n+1}e^{\sqrt3\pi n}
(\xi_n^-)^2\Im(\o\wh v_{n}),\qq\forall\qq n\in\N,
\]
where
\[
\lb{whgvxi}
\xi_n^\pm=1\pm(-1)^ne^{-\sqrt3\pi n}.
\]

ii) Let $(\l,\gu)\in\L_+\ts\cB_1(\ve)$ and let  $\wh p_0=\wh q_0=0$.
Then the function $\Delta$ satisfies
\[
\lb{estgamma}
\Big|\Delta(\l)-\Delta_0(\l)
-{i\b(\l)\/9\sqrt3z^5}\Big|\le {C|e^{z(\o^2+2)}|\|\gu\|_1^2\/|z|^6},
\]
for some $C>0$,
where $\Delta_0$ is given by \er{defphi0}.
\end{lemma}

\no {\bf Proof.}
i) The definition \er{defbeta} gives
$$
\begin{aligned}
\b=e^{(\o^2+2)z}\big(\zeta_{11}+\zeta_{22}+\zeta_{33}\big)
+e^{(\o+2\o^2)z}\big(\zeta_{21}+\zeta_{32}\big)
+e^{(1+2\o)z}\big(\zeta_{12}+\zeta_{23}\big)
-e^{(\o^2+2\o)z}\zeta_{22}
\\
-e^{(\o+2)z}\big(\zeta_{21}+\zeta_{12}+\zeta_{33}\big)
-e^{(1+2\o^2)z}\big(\zeta_{11}+\zeta_{32}+\zeta_{23}\big).
\end{aligned}
$$
The identities \er{o-o} yield
\[
\lb{betalambda}
\b=A_1+A_2+A_3+(e^{i\sqrt3\o^2z}-e^{-i\sqrt3\o z})\zeta_{33}
+(e^{i\sqrt3z}-e^{-i\sqrt3z})\zeta_{23},
\]
where
$$
\begin{aligned}
A_1=e^{i\sqrt3\o^2z}\big(\zeta_{11}+\zeta_{22}\big)
-e^{-i\sqrt3\o z}\big(\zeta_{21}+\zeta_{12}\big),\qq
A_2=-e^{-i\sqrt3\o^2z}\zeta_{22}
+e^{i\sqrt3\o z}\big(\zeta_{21}+\zeta_{32}\big),
\\
A_3=e^{i\sqrt3z}\zeta_{12}
-e^{-i\sqrt3z}\big(\zeta_{11}+\zeta_{32}\big).
\end{aligned}
$$
The identity \er{idzeta} gives
$$
A_1
=2i\Im(e^{i\sqrt3\o^2z}\theta_2)+2i\Im \theta_2
-2i\Im(e^{-i\sqrt3\o z}\theta_2)
-2i\Im(\o e^{-i\sqrt3\o z}\theta_3),
$$
$$
A_2
=2i\Im\big(e^{i\sqrt3\o z}\theta_2\big)
-2i\Im\big(\o e^{-i\sqrt3\o^2 z}\theta_3\big),
$$
$$
A_3
=-2i\Im\big(e^{-i\sqrt3z}(1+e^{-i\sqrt3\o^2z})\theta_2\big)
+2i\Im\big(\o (1+e^{i\sqrt3z})\theta_3\big),
$$
for $\l>0$.
The identity \er{betalambda} yields \er{idbeta}.

Let $\l=\m_{n}^o=z^3,z=z_n^o$.
Substituting the identities $i\sqrt3\o^2z=(\sqrt3-i)\pi n$,
$i\sqrt3\o z=-(\sqrt3+i)\pi n$, and $e^{i\sqrt3z}=e^{i2\pi n}=1$
into the identity \er{idbeta} we obtain
$$
\b(\m_{n}^o)=
2i\big(2-(-1)^ne^{\sqrt3\pi n}-(-1)^ne^{-\sqrt3\pi n}\big)
\Im\big(\o \theta_3(\m_{n}^o)\big),
$$
The definitions \er{whgvxi} imply
$$
2-(-1)^ne^{\sqrt3\pi n}-(-1)^ne^{-\sqrt3\pi n}=
(-1)^{n+1}e^{\sqrt3\pi n}(\xi_n^-)^2,
$$
which yields \er{betaatmn0}.

ii) Let $(\l,\gu)\in\L_+\ts\cB_1(\ve)$ and let  $\wh p_0=\wh q_0=0$.
The estimates \er{defwj} imply
\[
|w_{jk}(\l)-\zeta_{jk}(\l)|\le{ C\|\gu\|_1^2\/|z|},
\]
for all $j,k=1,2,3,\l\in\L_+$, and for some $C>0$.
The definitions \er{defbeta} and \er{defWW},
give
$$
|W(\l)-\b(\l)|
\le {C|e^{z(\o^2+2)}|\|\gu\|_1^2\/|z|},
$$
for some $C>0$. Then the identity \er{iddetphi234} implies  \er{estgamma}.~\BBox

\medskip

Introduce the function
$$
\g(z)=e^{-{3\/2}z}\b(z^3),
$$
and the domains
$$
\Pi_a=\{z\in\C:\Re z>a,|\Im z|<1\},\qq
\Pi_a^+=\{z\in\C_+:\Re z>a,0<\Im z<1\},\qq a\in\R.
$$

\begin{lemma}

Let $(\l,\gu)\in\L_+\ts\cB_1(\ve)$ and let $\wh p_0=\wh q_0=0$.
Then the function $\g(z)$ is analytic on the domain
$\Pi_1$ and satisfies
\[
\lb{estgammastrip}
|\g(z)|\le C\|\gu\|_1,\qq z\in\Pi_1,
\]
\[
\lb{betaatmn01}
\g(z_n^o)
=i2(-1)^{n+1}(\xi_n^-)^2
\Im(\o\wh v_{n}),\qq n\in \N,
\]
\[
\lb{estgamma'}
|\g'(z)|\le C\|\gu\|_1\qq\text{on}\qq [1,+\iy),
\]
for some $C>0$.
\end{lemma}

\no {\bf Proof.} The function $\b$ is analytic on $\C\sm (-\iy,0]$, then
$\g$ is analytic on the domain $\Pi_1$.
 The identity \er{betaatmn0} gives \er{betaatmn01}. Moreover,
we have the estimates $\Re(\pm i\sqrt3 \o^j z)\le{3\/2}\Re z$
in $\Pi_1^+$
for all $j=1,2,3$.
Then the estimates \er{zetaogr} and the definition \er{betalambda}
imply \er{estgammastrip} on $\Pi_1^+$. Moreover, the function $i\gamma$ is
real on $\R_+$, then the symmetry principle extends
the estimate \er{estgammastrip} onto $\Pi_1$.
Cauchy's formula yields \er{estgamma'}.~\BBox

\subsection{Rough eigenvalue asymptotics}

Introduce the analytic function
\[
\lb{defvk00}
\vk_0(z)=-i3\sqrt3e^{-{3\/2}z}z^3\Delta_0(z^3),\qq z\in\C\sm(-\iy,0],
\]
where $\Delta_0$ has the form \er{ascA0}.
Recall that $\m_n^o=(z_n^o)^3,n\in\Z\sm\{0\}$, are zeros of the
function $\Delta_0(\l)$, then
\[
\lb{vk0zo}
\vk_0(z_n^o)=0,\qq \forall \ \ n\in\N.
\]
Now we determine a rough asymptotics of the eigenvalues.

\begin{lemma}
i) The function $\vk_0$ satisfies
\[
\lb{vk00atzn0}
\vk_0'(z_n^o)=(-1)^{n+1}i\sqrt3(\xi_n^-)^2,
\qq \forall \ \ n\in\N,
\]
\[
\lb{estvk00''}
|\vk_0''(z)|\le C\qq\text{on}\qq [1,+\iy),
\]
for some $C>0$, where $\xi_n^-$ is given by \er{whgvxi}.

ii) Let $\gu\in\cB_1(\ve)$ and let  $\wh p_0=\wh q_0=0$.
Then
the eigenvalues $\m_n$ satisfy
\[
\lb{rest3pev1}
\Big|\m_n^{1\/3}-z_n^o\Big|\le
{C\|\gu\|_1\/n^2},
\]
for all $n\in\N$ and for some $C>0$.

\end{lemma}

\no {\bf Proof.}
i) The identities \er{ascA0} and \er{vk0zo} give
$$
\vk_0'(z_n^o)=-i3\sqrt3e^{-\sqrt3\pi n}{d(z^3\Delta_0(z^3))\/dz}\Big|_{z=z_n^o}
=-2i\sqrt3e^{-\sqrt3\pi n}\big((-1)^n\cosh (\sqrt3\pi n)-1\big),
$$
which yields \er{vk00atzn0}. The identity \er{ascA0} implies
$|\vk_0(z)|\le C$ on the strip
$\Pi_0$ for some $C>0$.
Cauchy's formula yields \er{estvk00''}.

ii) The estimate \er{estgamma} gives
\[
\lb{pertdetphi0}
\Delta
=\Delta_0+{e^{{3\/2}z}\a\/\l},\qq|\a(\l)|\le{C\|\gu\|_1\/|z|^2}.
\]
Introduce the function
$$
\vk(z)=-i3\sqrt3\l e^{-{3\/2}z}\Delta(z^3),\qq\l\in\Pi_1.
$$
Let $z_n=\m_n^{1\/3},n\in\N$.
Then $\vk(z_n)=0$ and, due to Lemma~\ref{Th3prami},
$z_n=z_n^o+\d_n\in\R,-1<\d_n<1$.
The identity \er{pertdetphi0} implies
$$
\vk(z_n)=\vk_0(z_n)+\a(\m_n)=\vk_0(z_n^o+\d_n)+\a(\m_n)
=\d_n\vk_0'(z_n^o)
+{\d_n^2\/2}\vk_0''(z)+\a(\m_n),
$$
where $z\in(z_n^o,z_n)$ and we used the identity $\vk_0(z_n^o)=0$.
The identity $\vk(z_n)=0$ gives
$$
\d_n=-{\vk_0'(z_n^o)\/2}\bigg(1-
\Big(1-{2\vk_0''(z)\a(\m_n)\/{\vk_0'}^2(z_n^o)}\Big)^{1\/2}\bigg).
$$
The estimates $|1-(1-\ve)^{1\/2}|\le|\ve|$ for all $\ve\in\R$ small enough
and the relations \er{estgamma}, \er{vk00atzn0}, and \er{estvk00''} give
$
|\d_n|\le{C\|\gu\|_1\/n^2},
$
for all $n\in\N$ and for some $C>0$, which yields \er{rest3pev1}.~\BBox

\subsection{Improvement of the eigenvalues asymptotics}
The definitions \er{deffn} imply
\[
\lb{idgagb}
\begin{aligned}
\ga_{-n}(\gu_*)=-{\wh p_{sn}'\/\sqrt3}-\wh q_{cn}=-\ga_n(\gu),\qq
\gb_{-n}(\gu_*)={\wh p_{cn}'\/\sqrt3}-\wh q_{sn}=\gb_n(\gu),
\\
\ga_{-n}(\gu_*^-)=-\ga_n(\gu^-)=-\ga_n(\gu),\qq
\gb_{-n}(\gu_*^-)=\gb_n(\gu^-)=-\gb_n(\gu).
\end{aligned}
\]
Now we determine asymptotics of the 3-point eigenvalues for the case
$\gu\in\cB_1(\ve)$ and $\wh p_0=\wh q_0=0$.
Recall that if $\gu\in\cB_1(\ve)$, then
$\m_n,n\in\N$, are zeros of the function $\phi_2$.

\begin{lemma}
\lb{Lm36}
Let $\gu\in\cB_1(\ve)$, let $\wh p_0=\wh q_0=0$, and
let $z_n=\m_n^{1\/3},\wt z_n=\wt\m_n^{1\/3},n\in\N$. Then
\[
\lb{aszn}
\max\Big\{\Big|z_n-z_n^o+{1\/(2\pi n)^2}\Big(\ga_n-{\gb_n\/\sqrt3}\Big)\Big|,
\Big|\wt z_n-z_n^o
+{1\/(2\pi n)^2}\Big(\ga_n+{\gb_n\/\sqrt3}\Big)\Big|\Big\}\le{C\|\gu\|_1^2\/n^3},
\]
for some $C>0$, where
$\ga_n,\gb_n$ are given by \er{deffn}.

\end{lemma}

\no {\bf Proof.}
Introduce the function
$$
\vk(z)=-i3\sqrt3z^3 e^{-{3\/2}z}\Delta(z^3),\qq  z\in\Pi_1.
$$
Let $\wh p_0=\wh q_0=0$ and
let $\l=\m_n,n\in\N,z_n=\l^{1\/3}$.
Then $\vk(z_n)=0$ and, due to the estimate \er{rest3pev1},
\[
\lb{aszrough}
z_n=z_n^o+\d_n,\qq\d_n\in\R,\qq |\d_n|\le{C\|\gu\|_1\/n^2},
\qq z_n^o={2\pi n\/\sqrt3}.
\]
The estimate \er{estgamma} implies
\[
\lb{estvkznr}
\vk(z_n)=\vk_0(z_n)+{\g(z_n)\/3z_n^2}+{\a_1(n)\/n^3}
=\vk_0(z_n)+{\g(z_n)\/(2\pi n)^2}
+{\a_2(n)\/n^3},\qq |\a_j(n)|\le C\|\gu\|_1^2,\qq j=1,2,
\]
where we used \er{estgammastrip}.
The relations \er{vk00atzn0} and \er{estvk00''}
and the estimate \er{aszrough} imply
\[
\lb{asvk00zn}
\vk_0(z_n)=\vk_0(z_n^o+\d_n)
=\d_n\vk_0'(z_n^o)+{\d_n^2\/2}\vk_0''(z)
=i(-1)^{n+1}\sqrt3(\xi_n^-)^2\d_n
+{\a_3(n)\/n^3},
\]
where $|\a_3(n)|\le C\|\gu\|_1^2$, $z\in(z_n^o,z_n)$,
$\xi_n^-$ are given by \er{whgvxi},
and we used the identity $\vk_0(z_n^o)=0$.
The relations \er{betaatmn01} and \er{estgamma'} imply
\[
\lb{asgammazn}
\g(z_n)=\g(z_n^o+\d_n)
=\g(z_n^o)+\d_n\g'(z)
=2i(-1)^{n+1}(\xi_n^-)^2\Im(\o\wh v_{n})
+{\a_4(n)\/n^2},
\]
where $|\a_4(n)|\le C\|\gu\|_1^2$, $z\in(z_n^o,z_n)$.
Substituting the asymptotics \er{asvk00zn} and \er{asgammazn}
into \er{estvkznr} we obtain
$$
\vk(z_n)
=-i(-1)^n\sqrt3(\xi_n^-)^2\d_n
-{2i(-1)^n\/(2\pi n)^2}(\xi_n^-)^2\Im(\o\wh v_{n})
+{\a_5(n)\/n^3},
$$
where $|\a_5(n)|\le C\|\gu\|_1^2$.
The identity $\vk(z_n)=0$ gives
$$
\d_n=-{\Im(\o\wh v_{n})\/2\sqrt3(\pi n)^2}+{\a_6(n)\/n^3},
$$
where $|\a_6(n)|\le C\|\gu\|_1^2$.
The identity
$$
\Im(\o\wh  v_{n})=
-{1\/2}\Im\Big((1-i\sqrt3)
\Big({i(\wh p_{cn}'-i\wh p_{sn}')\/\sqrt3}+\wh q_{cn}-i\wh q_{sn}\Big)\Big)=
{\sqrt3\/2}\Big(\ga_n-{\gb_n\/\sqrt3}\Big),
$$
gives
$$
\d_n=-{1\/(2\pi n)^2}\Big(\ga_n-{\gb_n\/\sqrt3}\Big)
+{\a_6(n)\/n^3},
$$
then \er{aszrough} gives the first estimate in \er{aszn}.
The identities \er{symev},  \er{idgagb}, and the first estimate in \er{aszn} give
$$
\wt z_n=-z_{-n}(\gu_*)=-z_{-n}^o
+{1\/(2\pi n)^2}\Big(\ga_{-n}(\gu_*)-{\gb_{-n}(\gu_*)\/\sqrt3}\Big)
+{\a_7(n)\/n^3}=z_{n}^o
-{1\/(2\pi n)^2}\Big(\ga_n+{\gb_n\/\sqrt3}\Big)+{\a_7(n)\/n^3},
$$
where $|\a_7(n)|\le C\|\gu\|_1^2$. This yields
the second estimate in \er{aszn}.~\BBox

\medskip

Lemma~\ref{Lm36} gives Theorem~\ref{Th3pram}.

\medskip

\no {\bf Proof of Theorem~\ref{Th3pram}.}
Due to Lemma~\ref{Th3prami},
there is exactly one simple real
eigenvalue $\m_n$ in each domain $\cD_n,n\in\Z_0=\Z\sm\{0\}$.
The first estimate in \er{aszn} yields \er{asmun} for $n>0$.
If $n<0$, then
the identities \er{symev} and  \er{idgagb} imply
$$
\m_{n}(\gu)=-\wt\m_{-n}(\gu_*)
=-\m_{-n}^o+\ga_{-n}(\gu_*)+{\gb_{-n}(\gu_*)\/\sqrt3}+{\a_8(n)\/n}
=\m_{n}^o-\ga_{n}(\gu)+{\gb_{n}(\gu)\/\sqrt3}+{\a_8(n)\/n},
$$
where $|\a_8(n)|\le C\|\gu\|_1^2$. This yields \er{asmun} for $n<0$.~\BBox

\section{Norming constants}
\setcounter{equation}{0}
\lb{Sectnc}

\subsection{Two representations of the monodromy matrix }
In this section we determine asymptotics of the norming constants defined
by \er{defnf}.
We first express the norming constants in terms of the multipliers
and the fundamental solutions. Then we determine the asymptotics
of the multipliers and the fundamental solutions.
This gives us the asymptotic expressions of the norming constants.

Introduce the monodromy matrix $M(\l),\l\in\C$, by
$$
M(\l)=\Phi(1,\l),
$$
where the fundamental matrix $\Phi$ is given by \er{deM}.
Now we obtain two representations
of the monodromy matrix as a product of bounded in $\L_+$ matrices
and a diagonal matrix, the first representation \er{factM}
 for the case $p,q\in L^2(\T)$ and the second representation \er{factM1}
 for the case
$p',q\in L^2(\T)$.
Assume $N=1$ in Lemmas~\ref{LmBest} and \ref{LmBest1}.

\begin{corollary}
i) Let $\gu\in\cB({1\/12})$.
Then the monodromy matrix satisfies
\[
\lb{factM}
M(\l)=\Psi_1(1,\l)\Psi_1^{-1}(0,\l)= \cU_1(\l)e^{z\O} G_1(\l)\cU_1^{-1}(\l),
\]
for all $\l\in\L_+$,
where $\Psi_1$ is given by  \er{defcA},
\[
\lb{defmUmF}
\cU_1(\l)= U_1(\l) X_1(1,\l),\qq
 G_1(\l)= X_1^{-1}(0,\l) X_1(1,\l),
\]
$ U_1$ is given by \er{defmaZ}, $ X_1$ is defined in Lemma~\ref{LmBest}~i),
the matrix-valued functions $\cU_1$ and $ G_1$ are analytic
in the domain $\L_+$, each of the functions $\cU_1(\l,\cdot)$
and $ G_1(\l,\cdot), \l\in\L_+$, is analytic on $\cB({1\/12})$.

ii) Let $(\l,\gu)\in\L_+\ts\cB_1(\ve)$ and let $\wh p_0=\wh q_0=0$.
Then
\[
\lb{factM1}
M(\l)=\Psi_2(1,\l)\Psi_2^{-1}(0,\l)
= \cU_2(\l)e^{z\O} G_2(\l)\cU_2^{-1}(\l),
\]
where  $\Psi_2$ is given by \er{idPhiss},
\[
\lb{defmU1mF1}
\cU_2(\l)= U_1(\l) U_2(0,\l) X_2(1,\l),\qq
 G_2(\l)= X_2^{-1}(0,\l) X_2(1,\l),
\]
$U_2$ is given by \er{defcU1}, $ X_2$ is defined in Lemma~\ref{LmBest1}~i).

\end{corollary}

\no {\bf Proof.}
The identity \er{factPhi1}  gives \er{factM}.
If $\wh p_0=\wh q_0=0$, then the definition \er{Phi1Theta}
implies $\int_0^1\Theta(s,\l)ds=\O$.
The identities \er{factPhi2} and $U_2(1,\l)=U_2(0,\l)$ give \er{factM1}.~\BBox

\subsection{Estimates of the monodromy matrix}
Below we need the following estimates for the monodromy matrix.

\begin{lemma}
i) Let  $(\l,\gu)\in\L_+\ts \cB({1\/12})$.
Then the matrix-valued function $ G_1$,
given by \er{defmUmF}, satisfies
\[
\lb{asFrou}
\big| G_1(\l)-\1_3\big|\le{24\|\gu\|\/|z|}.
\]

ii) Let $(\l,\gu)\in\L_+\ts\cB_1(\ve)$.
Then the matrix-valued function $ G_2$,
given by \er{defmU1mF1}, satisfies
\[
\lb{mFlj}
| G_2(\l)-\1_3|\le {C\|\gu\|_1\/|z|^2},
\]
\[
\lb{4g.asG1ipr}
\Big| G_2(\l)-\1_3-{L_1(\l)\/z^2}-{L_2(\l)\/z^{4}}\Big|
\le {C\|\gu\|_1^3\/|z|^{6}},
\]
for some $C>0$,
where
\[
\lb{cB12}
L_1(\l)=R_1(1,\l)-R_1(0,\l),
\]
\[
\lb{cB122}
L_2(\l)=R_2(1,\l)-R_2(0,\l)+R_1^2(0,\l)
-R_1(0,\l)R_1(1,\l),
\]
$R_1$ and $R_2$ are given by \er{defmRn}.
The diagonal entries of the matrix $L_1$
satisfy
\[
\lb{trcB1}
L_{1,jj}=0,\qq j=1,2,3.
\]

\end{lemma}

\no {\bf Proof.}
i) Let  $(\l,\gu)\in\L_+\ts \cB({1\/12})$. The definition \er{defmUmF} implies
$$
| G_1(\l)-\1_3|
\le| X_1^{-1}(0,\l)-\1_3|+| X_1(1,\l)-\1_3|
+| X_1^{-1}(0,\l)-\1_3|| X_1(1,\l)-\1_3|.
$$
The estimates \er{estcX} and \er{estcX-1} give \er{asFrou}.

ii) Let $(\l,\gu)\in\L_+\ts\cB_1(\ve)$. Using the definition \er{defmU1mF1}
and the estimates \er{estcXimp'} and repeating the previous arguments
we obtain \er{mFlj}.
Furthermore,
the estimates \er{estcXimp'} imply
$$
\begin{aligned}
&\max\Big\{\Big| X_2(1,\l)-\1_{3}-{R_1(1,\l)\/z^2}-{R_2(1,\l)\/z^{4}}\Big|,
\\
&\Big| X_2^{-1}(0,\l)-\1_{3}+{R_1(0,\l)\/z^2}
-{R_1^2(0,\l)-R_2(0,\l)\/z^{4}}\Big|\Big\}
\le {C\|\gu\|_1^3\/z^{6}}.
\end{aligned}
$$
These estimates and the definition \er{defmU1mF1}  yield \er{4g.asG1ipr},
where $L_1,L_2$ are given by
\er{cB12} and \er{cB122}.
The definitions \er{4g.dcLipr} and \er{Kljpr} and the identity \er{Phi1jj=0}
give
\[
\lb{cKjj=0}
R_{1,jj}(1,\l)=0,
\qq
R_{1,jj}(0,\l)=-\int_0^1 F_{2,jj}(s,\l)ds=0.
\]
The identity \er{cB12} yields \er{trcB1}.~\BBox

\subsection{Multipliers}
Due to \er{factM}, the multipliers $\t_1,\t_2$, and $\t_3$
are eigenvalues of the matrix
$e^{z\O} G_1=(e^{z\o^j} G_{1,jk})_{j,k=1}^3$.
Introduce Gershgorin's discs,
on the $\t$-plane
\[
\lb{Gershd}
\begin{aligned}
\Gamma_1=\{\t\in\C:|\t-e^{z\o} G_{1,11}|\le (| G_{1,12}|+| G_{1,13}|)|e^{z\o}|\},\\
\Gamma_2=\{\t\in\C:|\t-e^{z\o^2} G_{1,22}|\le (| G_{1,21}|+| G_{1,23}|)|e^{z\o^2}|\},\\
\Gamma_3=\{\t\in\C:|\t-e^{z} G_{1,33}|\le (| G_{1,31}|+| G_{1,32}|)|e^{z}|\}.
\end{aligned}
\]

\begin{lemma}
\lb{LmGD}
i) Let $\gu\in\cB(\ve)$.
Then
\[
\lb{Gd13}
\Gamma_1\cap\Gamma_3=\es,\qq\text{as}\qq \l\in\L_+,\qqq
\Gamma_2\cap\Gamma_3=\es,\qq\text{as}\qq \l\in\mD,
\]
where $\mD$ is given by \er{dmD}.

ii) Let  $(\l,\gu)\in\mD\ts\cB(\ve)$.
Then there exists exactly one simple multiplier $\t_3$ in the disc $\Gamma_3$
Moreover,
\[
\lb{lom}
|\t_3(\l)-e^{z}|\le {C\|\gu\|\/|z|}|e^{z}|,
\]
for some $C>0$.
The function $\t_3$
is analytic on the domain $\mD$.
Moreover, $\t_3(\l,\cdot),\l\in\mD$, is analytic on the ball
$\cB(\ve)$. Furthermore,
the function $\t_3$ is real for real $\l$ and satisfies
\[
\lb{t3>0}
\t_3(\l)>0,\qq \forall\ \ \l>1.
\]

\end{lemma}

\no {\bf Proof.}
i) Discs do not intersect if the distance
between their centers is greater than the sum of their radii.
Thus, the first relation in \er{Gd13} holds iff
$$
|e^{z\o} G_{1,11}-e^{z} G_{1,33}|>
(| G_{1,12}|+| G_{1,13}|)|e^{z\o}|+
(| G_{1,31}|+| G_{1,32}|)|e^{z}|,
$$
which is equivalent to
$$
r_1:=|e^{(\o-1)z} G_{1,11}- G_{1,33}|-
(| G_{1,12}|+| G_{1,13}|)|e^{(\o-1)z}|-
| G_{1,31}|-| G_{1,32}|>0.
$$
Let $(\l,\gu)\in\L_+\ts\cB({1\/12})$.
The estimate \er{asFrou} gives
$$
| G_{1,jj}|\ge 1-{24\|\gu\|\/|z|}\ge1-24\|\gu\|,\qq
| G_{1,jk}|\le{24\|\gu\|\/|z|}\le24\|\gu\|,
\qq j=1,2,3,\ \ j\ne k,
$$
and using the estimate \er{uest01-3} we obtain
$$
\begin{aligned}
|e^{(\o-1)z} G_{1,11}- G_{1,33}|\ge
| G_{1,33}|-|e^{(\o-1)z} G_{1,11}|\ge
1-24\|\gu\|-e^{-{3\/2} }(1-24\|\gu\|)
\\
=1-e^{-{3\/2} }-24\|\gu\|(1+e^{-{3\/2} }).
\end{aligned}
$$
Moreover,
$$
(| G_{1,12}|+| G_{1,13}|)|e^{(\o-1)z}|+
| G_{1,31}|+| G_{1,32}|\le48\|\gu\|(1+e^{-{3\/2} }),
$$
which yields
$$
r_1\ge
1-e^{-{3\/2} }-72\|\gu\|(1+e^{-{3\/2} }).
$$
If $\|\gu\|\le\ve$, then $r_1>0$, which yields the first relation in \er{Gd13}.

Furthermore, $\Gamma_2\cap\Gamma_3=\es$ iff
$$
|e^{z\o^2} G_{1,22}-e^{z} G_{1,33}|>
(| G_{1,21}|+| G_{1,23}|)|e^{z\o^2}|+(| G_{1,31}|+| G_{1,32}|)|e^{z}|,
$$
which is equivalent to
$$
r_2:=|e^{(\o^2-1)z} G_{1,22}- G_{1,33}|-
(| G_{1,21}|+| G_{1,23}|)|e^{(\o^2-1)z}|-| G_{1,31}|-| G_{1,32}|>0.
$$
Let $(\l,\gu)\in\L_+\ts\cB({1\/12})$.
The estimates \er{asFrou} and
$|e^{(\o^2-1)z}|\le 1$ give
$$
(| G_{1,21}|+| G_{1,23}|)|e^{(\o^2-1)z}|+| G_{1,31}|+| G_{1,32}|
\le{48\|\gu\|\/|z|}\le48\|\gu\|.
$$
Moreover,
$$
\begin{aligned}
|e^{(\o^2-1)z} G_{1,22}- G_{1,33}|
=|e^{(\o^2-1)z}-1+e^{(\o^2-1)z}( G_{1,22}-1)-( G_{1,33}-1)|
\\
\ge|e^{(\o^2-1)z}-1|-
|e^{(\o^2-1)z}|| G_{1,22}-1|-| G_{1,33}-1|
\ge|e^{(\o^2-1)z}-1|-24\|\gu\|.
\end{aligned}
$$
Thus
\[
\lb{est1}
r_2\ge|e^{(\o^2-1)z}-1|-72\|\gu\|.
\]
The estimates \er{est1} and \er{lesto2-3} give
$
r_2 \ge{1\/3}-72\|\gu\|.
$
If $\|\gu\|<\ve$, then
$
r_2>0,
$
which yields the second relation in \er{Gd13}.

ii) Let $(\l,\gu)\in\mD\ts\cB(\ve)$.
Then the disc $\Gamma_3$ is separated
from $\Gamma_1$ and $\Gamma_2$. Due to Gershgorin's theorem,
see \cite[Ch~6.1]{HJ85},
there exists exactly one simple multiplier $\t_3$ in the disc $\Gamma_3$.
The estimates
$$
|\t_3(\l)-e^{z}|\le|e^{z}|\big(| G_{1,31}|+| G_{1,32}|+| G_{1,33}-1|\big)
\le{C\|\gu\|\/|z|}|e^{z}|
$$
give \er{lom}.
The matrix-valued function $M$ is entire, then
the functions $\t_3$
is analytic on the domain $\mD$.
Due to Lemma~\ref{T21}, the function $M(\l,\cdot),\l\in\C$, is
analytic on $ \gH$, then each of the function
$\t_3(\l,\cdot),\l\in\mD$, is analytic on the ball
$\cB(\ve)$.
Moreover, if $\l$ is real,
then $\t_1=\ol\t_2$ or $\t_1$ and
$\t_2$ are real. The identity $\t_1\t_2\t_3=1$ provides
that $\t_3$ is also real.
The estimate \er{lom} gives
$$
\t_3(\l)\ge e^z- {C\|\gu\|\/z}e^{z},\qqq z>0,
$$
which yields \er{t3>0}.~\BBox

\subsection{Asymptotics of the multiplier $\t_3$}

We determine asymptotics of the multiplier $\t_3$ in the case $p',q\in L^2(\T)$.

\begin{lemma}

Let $(\l,\gu)\in\mD\ts\cB_1(\ve)$ and let $\wh p_o=\wh q_o=0$.
Then the multiplier $\t_3$ satisfies
\[
\lb{astau3}
|e^{-z}\t_3(\l)-1|\le {C\|\gu\|_1^2\/|z|^4}.
\]

\end{lemma}

\no {\bf Proof.}
The estimates \er{lom} show that the multiplier $\t_3$ has the form
\[
\lb{tau31}
\t_3=e^z(1+\d),
\]
where $|\d|\le{C\|\gu\|\/|z|}$.
Let $(\l,\gu)\in\L_+\ts\cB_1(\ve)$ and let $\wh p_o=\wh q_o=0$.
The identity \er{factM1} yields that $\t_3$ is an eigenvalue of the matrix
$e^{z\O} G_2$, where $G_2$ is given by \er{defmU1mF1}.
Due to \er{mFlj}, \er{4g.asG1ipr}, and \er{trcB1},
the matrix $ G_2$ has the form
\[
\lb{ezomf}
 G_2=\ma
1+{\alpha_{11}\/z^4}&{\alpha_{12}\/z^2}&{\alpha_{13}\/z^2}\\
{\alpha_{21}\/z^2}&1+{\alpha_{22}\/z^4}&{\alpha_{23}\/z^2}\\
{\alpha_{31}\/z^2}&{\alpha_{32}\/z^2}&1+{\alpha_{33}\/z^4}
\am,
\]
where
\[
\lb{alphajjjk}
|\alpha_{jj}(\l)|\le C\|\gu\|_1^2,\qq
|\alpha_{jk}(\l)|\le C\|\gu\|_1,\qq j,k=1,2,3.
\]
The identity \er{tau31} implies
$$
\det(e^{z\O} G_2-\t_3\1_3)=e^{3z}\det\big(e^{z(\O-\1_3)} G_2-(1+\d)\1_3\big)
=e^{3z}\det( G_3-\d_1\1_3),
$$
where
\[
\lb{tau32}
\d_1=\d-{\alpha_{33}\/z^4},
\]
$$
\begin{aligned}
&G_3=e^{z(\O-\1_3)} G_2-\Big(1+{\alpha_{33}\/z^4}\Big)\1_3=
\\
&\ma
e^{(\o-1)z}(1+{\alpha_{11}\/z^4})-1-{\alpha_{33}\/z^4}&e^{(\o^2-1)z}{\alpha_{12}\/z^2}&{\alpha_{13}\/z^2}\\
e^{(\o-1)z}{\alpha_{21}\/z^2}&e^{(\o^2-1)z}(1+{\alpha_{22}\/z^4})-1-{\alpha_{33}\/z^4}&{\alpha_{23}\/z^2}\\
e^{(\o-1)z}{\alpha_{31}\/z^2}&e^{(\o^2-1)z}{\alpha_{32}\/z^2}&0
\am.
\end{aligned}
$$
Then $\d_1$ is the eigenvalue of the matrix $ G_3$ and,
due to Gershgorin's theorem and \er{alphajjjk}, it satisfies the estimate
$$
|\d_1|\le {\max\{|\alpha_{13}|,|\alpha_{23}|\}\/|z|^2}\le {C\|\gu\|_1\/|z|^2},
$$
for some $C>0$.
Moreover,
$$
0=\det( G_3-\d_1\1_3)=-\d_1(e^{(\o-1)z}-1)(e^{(\o^2-1)z}-1)+{\alpha\/z^4},
$$
where $|\alpha(\l)|\le C\|\gu\|_1^2$,
for some $C>0$.
If $\l\in\mD$, then the estimates \er{uest01-3} and \er{lesto2-3} give
$|\d_1|\le{C\|\gu\|_1^2\/|z|^4}$,
for some $C>0$.
The identity \er{tau32} and the estimate
\er{alphajjjk} yield $|\d|\le{C\|\gu\|_1^2\/|z|^4}$,
then \er{tau31} implies \er{astau3}.~\BBox

\subsection{Eigenfunctions}
Introduce the function
\[
\lb{defgh}
\eta(t,\l)
=\det\ma\vp_2(t,\l)&\vp_3(t,\l)\\
\vp_2(1,\l)&\vp_3(1,\l)\am.
\]
Note that
\[
\lb{gh012}
\eta(0,\l)=\eta(1,\l)=0,\qq \eta(2,\l)=-\Delta(\l),
\]
where $\Delta$ is given by \er{defsi}.

\begin{lemma}
i) Let $\gu\in \gH$ and let
$\mu\in\C$ be an eigenvalue
of the 3-point Dirichlet problem for Eq.~\er{1b}.
Then the function $\eta(t,\m),t\in[0,2]$, given by \er{defgh},
is an eigenfunction corresponding to the eigenvalue $\mu$.
Moreover, $\eta'(0,\mu)\ne 0$ and
\[
\lb{ef'1}
{\eta'(1,\mu )\/\eta'(0,\mu  )}=
-{\wt\vp_3(1,\mu )\/\vp_3(1,\mu )},
\]
for all $t\in[0,2]$.

ii) Let $\gu\in\cB_1(\ve)$. Then the norming constants, given by \er{defnf}, satisfy
\[
\lb{idncr}
h_{sn}=8(\pi n)^2\log\Big((-1)^{n+1}{\vp_3(1,\wt\m_n)\/\wt\vp_3(1,\wt\m_n)}\t_3^{-{1\/2}}(\wt\m_n)\Big),
\qq \forall\qq n\in\N.
\]
\end{lemma}

\no {\bf Proof.}
i) Recall that $\mu  $ is a zero of the entire function
$\Delta$.
The definition \er{defgh} shows that $\eta$ satisfies Eq.~\er{1b} and
the identities \er{gh012} yield $\eta(0,\mu)=\eta(1,\mu)=\eta(2,\mu)=0$.
Therefore, $\eta$ is an eigenfunction and $\eta'(0,\mu)\ne 0$.
The identity \er{simMt} implies
$$
{\eta'(1,\mu)\/\eta'(0,\mu)}={\det\ma\vp_2'(1,\mu)&\vp_3'(1,\mu)\\
\vp_2(1,\mu)&\vp_3(1,\mu)\am\/\det\ma\vp_2'(0,\mu)&\vp_3'(0,\mu)\\
\vp_2(1,\mu)&\vp_3(1,\mu)\am}=-{\wt\vp_3(1,\mu)\/\vp_3(1,\mu)},
$$
which yields  \er{ef'1}.

ii) The definition \er{defnf} yields
$$
h_{sn}
=8(\pi n)^2\log\Big((-1)^n{\wt\eta'(1,\wt\m_n)\/\wt\eta'(0,\wt\m_n)}\t_3^{-{1\/2}}(\wt\m_n)\Big)
$$
for all $n\in\N$. The identity \er{ef'1} gives \er{idncr}.~\BBox

\subsection{Sharp asymptotics of the fundamental solution}
In the unperturbed case  $\gu=0$ the fundamental solutions have the form
\[
\lb{unpfs}
\vp_j^0={1\/3z^{j-1}}\sum_{n=1}^3\o^{n(1-j)}e^{\o^n zx},\qq j=1,2,3.
\]

\begin{lemma} Let $\gu\in\cB_1(\ve)$ and let $\wh p_0=\wh q_0=0$.
Then

i) The fundamental solution $\vp_3$ to Eq.~\er{1b}
satisfies
\[
\lb{estvp1at1}
\Big|e^{-z}\Big(\vp_3(1,\l)-\vp_3^0(1,\l)
-{2p(0)\/3\l}\vp_2^0(1,\l)-{b(\l)\/9z^4}\Big)\Big|
\le {C\|\gu\|_1^2\/|z|^6},
\]
for all $\l\in\Pi_1$ and for some $C>0$, where
\[
\lb{idBatvp3}
b=-e^{\o z}\theta_3-e^{\o^2z}\theta_6
-\o e^{\o z}\theta_4-\o e^{z}\theta_2
-\o^2e^{\o^2 z}\theta_1-\o^2e^{z}\theta_5,
\]
$\theta_j$ are given by \er{a145}.
Moreover, if $n\in\N$, then
\[
\lb{Batev}
b(\m_{n}^o)=2e^{z_n^o}\Re b_n,\qq
b_n=(-1)^{n+1}e^{-\sqrt3\pi n}\wh  v_n
-\o  v_n^--\o^2 (-1)^n e^{-\sqrt3\pi n }v_n^+,
\]
where
$\wh  v_n, v_ n^-, v_n^+$ are given by \er{whgvin+}.

ii) The fundamental solution $\wt\vp_3$ to Eq.~\er{1btr}
satisfies
\[
\lb{aswtvp3}
\Big|e^{\o z}\Big(\wt\vp_3(1,\l)-\wt\vp_3^0(1,\l)
+{2p(0)\/3\l}\wt\vp_2^0(1,\l)+{\wt b(\l)\/9z^4}\Big)\Big|\le{C\|\gu\|_1^2\/|z|^6},
\]
for all $\l\in\Pi_1$ and for some $C>0$, where
\[
\lb{idwtB}
\wt b=-e^{-\o^2z}\theta_3-e^{-\o z}\theta_6
-\o e^{-z}\theta_4-\o e^{-\o z}\theta_2
-\o^2e^{- z}\theta_1-\o^2e^{-\o^2 z}\theta_5.
\]
Moreover, if $n\in\N$, then
\[
\lb{wtBatzn0}
\wt b(\m_{n}^o)=2(-1)^{n+1}e^{{\pi n\/\sqrt3}}\Re \wt b_n,\qq
\wt b_n=\wh  v_{n}+\o  v_n^-+\o^2 (-1)^n e^{-\sqrt3\pi n }v_n^+.
\]

\end{lemma}

\no {\bf Proof.}
Let $\gu\in\cB_1(\ve)$, $\wh p_0=\wh q_0=0$, and let $\l\in\L_+$.
 The identities \er{idPhiss} and \er{factM1} give
\[
\lb{MitoS}
M(\l)= U_1(\l) \cS(1,\l)
e^{z\O}\cS^{-1}(0,\l) U_1^{-1}(\l),\qq \cS(x,\l)=U_2(x,\l) X_2(x,\l).
\]
This identity implies
\[
\lb{vp3intomm}
\vp_3(1,\l)=M_{13}(\l)
={1\/3z^2}\sum_{j,k,l=1}^3\o^le^{z\o^k}\cS_{jk}(1,\l)(\cS^{-1})_{kl}(0,\l),
\]
where we used the identities
$ U_{1,1j}(\l)=1$, $( U_1^{-1})_{l3}(\l)={\o^l\/3z^2}$, see \er{defmaZ}.
The definition \er{defcU1} and the estimates \er{estcXimp'}
and \er{estKI3}  imply
\[
\lb{ascS}
\cS(x,\l)=\Big(\1_3+{p(x) \O_3\/3z^2}\Big)
\Big(\1_3+{1\/3z^2}\int_0^1\cN(x,s,\l)ds+{\a_1(x,\l)\/z^4}\Big)
=\1_3+{\cW(x,\l)\/3z^2}+{\a_2(x,\l)\/z^4},
\]
\[
\lb{ascS-1}
\cS^{-1}(x,\l)=\1_3-{\cW(x,\l)\/3z^2}+{\a_3(x,\l)\/z^4},
\]
where $\cN$ is given by \er{defL0},
$\sup_{x\in[0,1]}|\a_j(x,\l)|\le C\|\gu\|_1^2,j=1,2,3$, and
\[
\lb{defcW}
\cW(x,\l)=p(x) \O_3+\int_0^1\cN(x,s,\l)ds.
\]
These asymptotics yield
$$
\cS_{jk}(1,\l)(\cS^{-1})_{kl}(0,\l)
=\d_{jk}\d_{kl}+{\d_{kl}\cW_{jk}(1,\l)-\d_{jk}\cW_{kl}(0,\l)\/3z^2}
+{\a_6(\l)\/z^4}
$$
where $|\a_6(\l)|\le C\|\gu\|_1^2$.
Substituting these asymptotics into \er{vp3intomm}
and using \er{unpfs} and  the estimates
$\Re(\o^k-1)\le 0,k=1,2,3$, we obtain
\[
\lb{asvp3at1pr}
\vp_3(1,\l)
=\vp_3^0(1,\l)+{1\/9z^4}
\Big(\sum_{j,l=1}^3e^{z\o^j}W_{jl}(\l)
+{e^{z}\a_7(\l)\/z^2}\Big),\qq|\a_7(\l)|\le C\|\gu\|_1^2,
\]
where
$$
W(\l)=\O\cW^\top(1,\l)-\cW(0,\l)\O=p(0)A+B(\l),
$$
\[
\lb{G-Ga}
A=\O \O_3^\top- \O_3\O={2i\/\sqrt3}\ma
0&-1&\o\\
1&0&-\o^2\\
-\o&\o^2&0
\am,
\]
\[
\lb{G-G}
B(\l)=\int_0^1\big(\O \cN^\top(1,s,\l)-\cN(0,s,\l)\O\big)ds,
\]
where $\cN$ is given by \er{defL0},
here we used \er{defcW} and  \er{defW1ss}.
The definition \er{G-G} gives
\[
\lb{BjloG}
B_{jl}(\l)=\int_0^1\big(\o^j \cN_{lj}(1,s,\l)-\o^l\cN_{jl}(0,s,\l)\big)ds,
\qq l,j=1,2,3.
\]
The identity \er{idL0} implies
\[
\lb{L0lj=0}
\cN_{jl}(0,s,\l)=\cN_{lj}(1,s,\l)=0,\qq 0\le j\le l\le 3,\qq s\in[0,1].
\]
Then  $B_{jl}=0,0\le j\le l\le 3$.
The identities \er{idL0}, \er{a145}, and \er{BjloG} give
\[
\lb{bjl}
B_{21}
=-e^{i\sqrt3z}\theta_3-\theta_6,
\qq
B_{31}
=-\o e^{-i\sqrt3\o^2z}\theta_4-\o\theta_2,
\qq
B_{32}
=-\o^2e^{i\sqrt3\o z}\theta_1-\o^2\theta_5.
\]
Then the identity \er{asvp3at1pr} implies
\[
\lb{asvp3at1}
\vp_3(1,\l)=\vp_3^0(1,\l)+{1\/9z^4}\Big(p(0)a(\l)+b(\l)
+{e^{z}\a_7(\l)\/z^2}\Big),
\]
where
\[
\lb{idalbl}
a=\sum_{j,l=1}^3e^{\o^jz}A_{jl},
\qq
b=\sum_{j,l=1}^3e^{\o^jz}B_{jl}.
\]
The identities \er{G-Ga}, \er{o-o},  and \er{unpfs}
yield
$$
a(\l)
={2i\/\sqrt3}\Big((1-\o^2)e^{z\o^2}+(\o -1)e^{z\o}
-(\o -\o^2)e^{z}\Big)
=2(\o^2e^{z\o}+\o e^{z\o^2}+e^{z})
=6z\vp_2^0(1,\l).
$$
Substituting this identity into \er{asvp3at1} we obtain \er{estvp1at1}
for $\l\in\Pi_1^+$.
The identity \er{idalbl} yields
$$
b=e^{\o^2z}B_{21}+e^{z}\big(B_{31}+B_{32}\big).
$$
The identities \er{bjl} and \er{o-o} give
\er{idBatvp3}.

If $\l>0$, then we have
$\theta_2=\ol \theta_5,\theta_6=\ol \theta_3,\theta_1=\ol \theta_4$.
The identity \er{idBatvp3} gives
\[
\lb{brelal}
b=-2\Re(e^{\o z}\theta_3+\o^2e^{\o^2 z}\theta_1+\o e^{z}\theta_2).
\]
The function $b$ is analytic on $\C\sm (-\iy,0]$
and real on $\R_+$, then the symmetry principle extends
the estimate \er{estvp1at1} onto $\Pi_1$.

Substituting the identities \er{thetamn0} and the identities
$e^{\o z_n^o}=e^{\o^2z_n^o}=(-1)^ne^{-{\pi n\/\sqrt3}}$ into \er{brelal}
we obtain \er{Batev}.

ii) The identities \er{simMt} and  \er{MitoS} imply
$$
\wt M(\l) =J(M^{-1}(\l))^\top J
=J( U_1^{-1}(\l))^\top(\cS^{-1}(1,\l))^\top e^{-\O z}\cS^\top(0,\l) U_1^\top(\l)J.
$$
This yields
$$
\wt\vp_3(1,\l)=\wt M_{13}(\l)=
{1\/3z^2}\sum_{j,k,l=1}^3\o^je^{-\o^k z}(\cS^{-1}(1,\l))_{kj}
\cS_{lk}(0,\l),
$$
where we used the identities $J_{jk}=(-1)^{j+1}\d_{j,4-k}$,
$ U_{1,1j}(\l)=1$, $( U_1^{-1})_{l3}(\l)={\o^l\/3z^2}$, see
\er{simMt} and \er{defmaZ}.
The asymptotics \er{ascS} and \er{ascS-1} give
$$
\begin{aligned}
\wt\vp_3(1,\l)=\wt\vp_3^0(1,\l)+
{1\/9z^4}\sum_{j,l=1}^3e^{-\o^j z}
\Big(\o^j\cW_{lj}(0,\l)-\o^l\cW_{jl}(1,\l)+{\a_8(\l)\/z^2}\Big)
\\
=\wt\vp_3^0(1,\l)-
{1\/9z^4}\sum_{j,l=1}^3e^{-\o^j z}
\Big(W_{lj}(\l)+{\a_8(\l)\/z^2}\Big)
\end{aligned}
$$
where $|\a_8(\l)|\le C\|\gu\|_1^2$.
The definition \er{defcW} and the estimates
$\Re (-z)\le\Re(-\o^2 z)\le\Re (-\o z)$ imply
\[
\lb{aswtvp3pr}
\wt\vp_3(1,\l)=\wt\vp_3^0(1,\l)-
{1\/9z^4}\Big(p(0)\wt a(\l)+\wt b(\l)+{e^{-\o z}\a_9(\l)\/z^2}\Big),
\]
where $|\a_9(\l)|\le C\|\gu\|_1^2$,
$$
\wt a=\sum_{j=1}^3e^{-\o^j z}\sum_{l=1}^3A_{lj},\qq
\wt b=\sum_{j,l=1}^3e^{-\o^j z}B_{lj}.
$$
The identity \er{G-Ga} yields
$$
\wt a(\l)
={2i\/\sqrt3}\big((1-\o)e^{-\o z}+(\o^2-1)e^{-\o^2 z}+(\o-\o^2)e^{-z}\big)
=-2\big(\o^2e^{-\o z}+\o e^{-\o^2 z}+e^{-z}\big)=6z\wt\vp_2^0(1,\l),
$$
where we used the identities \er{o-o} and the definitions \er{unpfs}.
Then the asymptotics \er{aswtvp3pr} gives \er{aswtvp3} for $\l\in\Pi_1^+$.
The identities \er{L0lj=0} yield
$$
\wt b=e^{-\o z}\big(B_{21}+B_{31}\big)+e^{-\o^2 z}B_{32}.
$$
The identities \er{bjl} and \er{o-o} give
\er{idwtB}.

If $\l>0$, then we have
$\theta_2=\ol \theta_5,\theta_6=\ol \theta_3,\theta_1=\ol \theta_4$.
The identity \er{idwtB} implies
$$
\wt b=-2\Re(e^{-\o^2z}\theta_3+\o^2e^{- z}\theta_1+\o e^{-\o z}\theta_2).
$$
The function $\wt b$ is analytic on $\C\sm (-\iy,0]$
and real on $\R_+$, then the symmetry principle extends
the estimate \er{aswtvp3} onto $\Pi_1$.
The identities \er{thetamn0}
and $e^{-\o z_n^o}=e^{-\o^2 z_n^o}=(-1)^{n}e^{{\pi n\/\sqrt3}}$
give \er{wtBatzn0}.~\BBox

\subsection{Additional asymptotics}
In order to determine asymptotics of the norming constants
we need the following results.

\begin{lemma} Let $\gu\in\cB_1(\ve)$ and let $\wh p_0=\wh q_0=0$. Then
\[
\lb{asvp3divwtvp3}
\Big|(-1)^{n+1}e^{-{\pi n\/\sqrt 3}}{\vp_3(1,\wt\m_n)\/\wt\vp_3(1,\wt\m_n)}
-1+{\gb_n\/2\sqrt3(\pi n)^2}\Big|
\le{C\|\gu\|_1^2\/n^3},
\]
for all $n\in\N$ and for some $C>0$, where $\gb_n$ is given by \er{deffn}.
\end{lemma}

\no {\bf Proof.}
Let $\gu\in\cB_1(\ve)$ and let
 $\wh p_0=\wh q_0=0$.
The estimate \er{estvp1at1} implies
\[
\lb{asvp3pr}
\vp_3(1,\l)=\vp_3^0(1,\l)
+{2p(0)\/3\l}\vp_2^0(1,\l)+{b(\l)\/9z^4}+e^{z}{\a_1(\l)\/z^6},\qq
|\a_1(\l)|\le C\|\gu\|_1^2,
\]
for all $\l\in\L_+$ and for some $C>0$.
Let $\l=\wt\m_n$ for some $n\in\N$ and let $\wt z_n=\l^{1\/3}$.
The estimate \er{rest3pev1} gives
\[
\lb{wtznpert}
\wt z_n=z_n^o+{\a_2(n)\/n^2},\qq z_n^o={2\pi n\/\sqrt3},
\qq |\a_2(n)|\le C\|\gu\|_1,
\]
for some $C>0$. The standard arguments yield
$$
\vp_3^0(1,\l)=\vp_3^0(1,\m_n^o)+{d\vp_3^0(1,z^3)\/dz}\Big|_{z=z_n^o}(\wt z_n-z_n^o)
+e^{z_n^o}{\a_3(n)\/n^6},\qq |\a_3(n)|\le C\|\gu\|_1^2.
$$
The identities
$$
\vp_3^0(1,\m_n^o)={e^{z_n^o}\xi_n^-\/(2\pi n)^2},\qq
{d\vp_3^0(1,z^3)\/dz}\Big|_{z=z_n^o}
={e^{z_n^o}\xi_n^-\/(2\pi n)^2}\Big(1-{\sqrt3\/\pi n}\Big),
$$
where
$\xi_n^-$ is given by \er{whgvxi},
imply
\[
\lb{asvp30}
\vp_3^0(1,\l)={e^{z_n^o}\xi_n^-\/(2\pi n)^2}
+{e^{z_n^o}\xi_n^-\/(2\pi n)^2}\Big(1-{\sqrt3\/\pi n}\Big)(\wt z_n-z_n^o)
+e^{z_n^o}{\a_3(n)\/n^6}.
\]
Moreover,
$$
\vp_2^0(1,\l)=\vp_2^0(1,\m_n^o)+e^{z_n^o}{\a_4(n)\/n^3},
\qq |\a_4(n)|\le C\|\gu\|_1.
$$
The identity
$$
\vp_2^0(1,\m_n^o)={\xi_n^-e^{z_n^o}\/2\sqrt 3\pi n}
$$
implies
\[
\lb{asvp20}
\vp_2^0(1,\l)={\xi_n^-e^{z_n^o}\/2\sqrt 3\pi n}
+e^{z_n^o}{\a_4(n)\/n^3}.
\]
Furthermore, the definition \er{idBatvp3} gives
\[
\lb{asBl}
b(\l)=b(\m_n^o)+e^{z_n^o}{\a_5(n)\/n^2},\qq|\a_5(n)|\le C\|\gu\|_1^2.
\]
Substituting the asymptotics \er{asvp30}, \er{asvp20}, and \er{asBl}
into \er{asvp3pr}, and using \er{Batev},
 we obtain
\[
\vp_3(1,\l)={e^{z_n^o}\xi_n^-\/(2\pi n)^2}
+{e^{z_n^o}\xi_n^-\/(2\pi n)^2}\Big(1-{\sqrt3\/\pi n}\Big)(\wt z_n-z_n^o)
+{e^{z_n^o}\xi_n^-p(0)\/8(\pi n)^4}
+{2e^{z_n^o}\Re b_n\/(2\pi n)^4}
+e^{z_n^o}{\a_6(n)\/n^6},
\]
where $|\a_6(n)|\le C\|\gu\|_1^2$, for some $C>0$.
The estimate \er{aszn} implies
\[
\lb{asz-zn0}
\wt z_n=z_n^o-{1\/(2\pi n)^2}\Big(\ga_n+{\gb_n\/\sqrt3}\Big)+{\a_7(n)\/n^3},
\]
where $|\a_7(n)|\le C\|\gu\|_1^2$ and $\ga_n,\gb_n$ are given by \er{deffn}.
Then
\[
\lb{asvp3wtmn}
\vp_3(1,\l)={\xi_n^-e^{z_n^o}\/(2\pi n)^2}\lt(1
-{1\/(2\pi n)^2}\Big(1-{\sqrt3\/\pi n}\Big)\Big(\ga_n+{\gb_n\/\sqrt3}\Big)
+{\Re b_n\/2(\pi n)^2\xi_n^-}+{p(0)\/2(\pi n)^2}
+{\a_8(n)\/n^3}\rt),
\]
where $|\a_8(n)|\le C\|\gu\|_1^2$, for some $C>0$.

The estimate \er{aswtvp3} implies
\[
\lb{aswtvp3wtmn}
\wt\vp_3(1,\l)=\wt\vp_3^0(1,\l)
-{2p(0)\/3\l}\wt\vp_2^0(1,\l)+{\wt b(\l)\/9z^4}+{e^{-\o z}\wt\a_1(\l)\/z^6},
\]
where  $|\wt\a_1(\l)|\le C\|\gu\|_1^2$, for all $\l\in\L_+$ and for some $C>0$.
Let $\l=\wt\m_n$ for some $n\in\N$ and let $\wt z_n=\l^{1\/3}$.
The estimate \er{wtznpert} yields
$$
\wt\vp_3^0(1,\l)=\wt\vp_3^0(1,\m_n^o)
+{d\wt\vp_3^0(1,z^3)\/dz}\Big|_{z=z_n^o}(\wt z_n-z_n^o)
+e^{\pi n\/\sqrt3}{\wt\a_2(n)\/n^6},\qq |\wt\a_2(n)|\le C\|\gu\|_1^2.
$$
The identities
$$
\wt\vp_3^0(1,\m_n^o)={(-1)^{n+1}e^{\pi n\/\sqrt3}\xi_n^-\/(2\pi n)^2},
\qq
{d\wt\vp_3^0(1,z^3)\/dz}\Big|_{z=z_n^o}
={(-1)^ne^{\pi n\/\sqrt3}\xi_n^-\/(2\pi n)^2}\Big(1+{\sqrt3\/\pi n}\Big),
$$
see \er{whgvxi}, give
\[
\lb{aswtvp30wtmn0}
\wt\vp_3^0(1,\l)={(-1)^{n+1}e^{\pi n\/\sqrt3}\xi_n^-\/(2\pi n)^2}
+{(-1)^ne^{\pi n\/\sqrt3}\xi_n^-\/(2\pi n)^2}\Big(1+{\sqrt3\/\pi n}\Big)(\wt z_n-z_n^o)
+e^{\pi n\/\sqrt3}{\wt\a_2(n)\/n^6}.
\]
Moreover,
\[
\lb{aswtvp20}
\wt\vp_2^0(1,\l)=\wt\vp_2^0(1,\m_n^o)+e^{\pi n\/\sqrt3}{\wt\a_3(n)\/n^3}
={(-1)^ne^{\pi n\/\sqrt3}\xi_n^-\/2\sqrt3\pi n}
+e^{\pi n\/\sqrt3}{\wt\a_3(n)\/n^3},
\qq |\wt\a_3(n)|\le C\|\gu\|_1.
\]
Furthermore, the definition \er{idwtB} and the identity \er{wtBatzn0}
 give
\[
\lb{aswtBl}
\wt b(\l)=\wt b(\m_n^o)+e^{\pi n\/\sqrt3}{\wt\a_4(n)\/n^2}
=2(-1)^ne^{{\pi n\/\sqrt3}}\Re \wt b_n+e^{\pi n\/\sqrt3}{\wt\a_4(n)\/n^2},
\qq|\wt\a_4(n)|\le C\|\gu\|_1^2.
\]
Substituting the asymptotics \er{aswtvp30wtmn0}, \er{aswtvp20},
\er{aswtBl}, and \er{asz-zn0} into \er{aswtvp3wtmn} we obtain
\[
\lb{aswtvp3wtmn1}
\wt\vp_3(1,\l)={(-1)^{n+1}\xi_n^-e^{\pi n\/\sqrt3}\/(2\pi n)^2}\Big(1
+{1\/(2\pi n)^2}\Big(1+{\sqrt3\/\pi n}\Big)\Big(\ga_n+{\gb_n\/\sqrt3}\Big)
+{p(0)\/2(\pi n)^2}
-{2\Re \wt b_n\/(2\pi n)^2{\xi_n^-}}+{\wt\a_5(n)\/n^3}\Big),
\]
where $|\wt\a_5(n)|\le C\|\gu\|_1^2$.

The asymptotics \er{asvp3wtmn} and \er{aswtvp3wtmn1} give
\[
\lb{vp3divwtvp3pr}
(-1)^{n+1}{\vp_3(1,\wt\m_n)\/\wt\vp_3(1,\wt\m_n)}
=e^{\pi n\/\sqrt3}\lt(1-{1\/2(\pi n)^2}\Big(\ga_n+{\gb_n\/\sqrt3}\Big)
+{\Re (b_n+\wt b_n)\/2(\pi n)^2{\xi_n^-}}
+{\wt\a_6(n)\/n^3}\rt),
\]
where $|\wt\a_6(n)|\le C\|\gu\|_1^2$.
The definitions \er{Batev} and \er{wtBatzn0} imply
$$
\Re(b_n+\wt b_n)=\xi_n^-\Re\wh  v_{n}=\xi_n^-\Re\Big({i\wh p_n'\/\sqrt3}+\wh q_n\Big)
=\xi_n^-\ga_n.
$$
Substituting this identity
into \er{vp3divwtvp3pr} we obtain
$$
(-1)^{n+1}{\vp_3(1,\wt\m_n)\/\wt\vp_3(1,\wt\m_n)}
=e^{\pi n\/\sqrt3}\lt(1-{\gb_n\/2\sqrt3(\pi n)^2}
+{\wt\a_6(n)\/n^3}\rt),
$$
which yields \er{asvp3divwtvp3}.~\BBox

\subsection{Asymptotics of the norming constants}

We are ready to prove Theorem~\ref{Thnf}.

\medskip

\no {\bf Proof of Theorem~\ref{Thnf}.}
Let $\l=\wt\m_n$ for some $n\in\N$ and let $\wt z_n=\l^{1\/3}$.
The estimate \er{aszn} gives
$$
\wt z_n=z_n^o-{1\/(2\pi n)^2}\Big(\ga_n+{\gb_n\/\sqrt3}\Big)+{\a_1(n)\/n^3},
$$
where $|\a_1(n)|\le C\|\gu\|_1^2$
for some $C>0$.
Then the estimate \er{astau3} gives
$$
\t_3(\wt\m_n)=e^{\wt z_n}\Big(1+{\a_2(n)\/n^4}\Big)
=e^{z_n^o}\Big(1-{\ga_n+{\gb_n\/\sqrt3}\/(2\pi n)^2}+{\a_3(n)\/n^3}\Big),
$$
where $|\a_j(n)|\le C\|\gu\|_1^2,j=2,3$.
This yields
$$
\t_3^{-{1\/2}}(\wt\m_n)
=e^{-{\pi n\/\sqrt 3}}\Big(1+{\ga_n+{\gb_n\/\sqrt3}\/2(2\pi n)^2}
+{\a_4(n)\/n^3}\Big),
$$
where $|\a_4(n)|\le C\|\gu\|_1^2$.
The estimate \er{asvp3divwtvp3} gives
$$
(-1)^{n+1}{\vp_3(1,\wt\m_n)\/\wt\vp_3(1,\wt\m_n)}
={e^{\pi n\/\sqrt 3}}\lt(1
-{\gb_n\/2\sqrt3(\pi n)^2}
+{\a_{5}(n)\/n^3}\rt),
\qq
|\a_5(n)|\le C\|\gu\|_1^2.
$$
These asymptotics imply
$$
(-1)^{n+1}{\vp_3(1,\wt\m_n)\/\wt\vp_3(1,\wt\m_n)}\t_3^{-{1\/2}}(\wt\m_n)
=1-{\gb_n\/2\sqrt3(\pi n)^2}+{\ga_n+{\gb_n\/\sqrt3}\/2(2\pi n)^2}
+{\a_6(n)\/n^4}=1+{\ga_n-\sqrt3\gb_n\/2(2\pi n)^2}+{\a_6(n)\/n^3},
$$
where $|\a_6(n)|\le C\|\gu\|_1^2$.
Substituting this asymptotics into \er{idncr} we obtain
$$
h_{sn}
=8(\pi n)^2\log\lt(1+{\ga_n-\sqrt3\gb_n\/2(2\pi n)^2}+{\a_6(n)\/n^3}\rt),
$$
which yields \er{asncr}.~\BBox

\medskip

In accordance with the identity \er{symev},
we introduce the norming constants $h_{s,-n},n\in\N$ by
\[
h_{s,-n}(\gu)=-h_{sn}(\gu_*^-),\qq n\in\N.
\]

\begin{corollary}
\lb{Cornf}
Let $\gu\in\cB_1(\ve)$ and let $\wh p_0=\wh q_0=0$. Then the norming
constants $h_{sn}$ satisfy the estimate
\er{asncr}
for all $n\in\Z_0$ and for some $C>0$.
\end{corollary}

\no {\bf Proof.}
Theorem~\ref{Thnf} gives \er{asncr} for $n>0$.
If $n<0$, then
the identities \er{symev} and \er{idgagb}
 imply
$$
h_{sn}(\gu)=-h_{s,-n}(\gu_*^-)
=-\ga_{-n}(\gu_*^-)+\sqrt3\gb_{-n}(\gu_*^-)+{\a_1(n)\/n^2}
=\ga_n(\gu)-\sqrt3\gb_n(\gu)+{\a(n)\/n},
$$
where $|\a(n)|\le C\|\gu\|_1^2$. This yields \er{asmun} for $n<0$.~\BBox

\bigskip

\no\small {\bf Acknowledgments.}
The authors  were supported by the RSF grant number 23-21-00023.


\begin{thebibliography}{9999}
\setlength{\itemsep}{-\parskip}
\footnotesize

\bibitem{A99} L. Amour, Determination of a third-order
operator from two of its spectra.
SIAM J. Math. Anal. 30 (1999) 1010--1028.

\bibitem{A01} L. Amour, Isospectral flows of third order operators.
SIAM J. Math. Anal. 32 (2001) 1375--1389.

\bibitem{BK11} A.Badanin,  E.Korotyaev.
Even order periodic operator on the real line.
Int. Math. Res. Not.,  rnr057 (2011) 53 p.

\bibitem{BK12} A.Badanin, E.Korotyaev. Spectral asymptotics for
the third order operator with periodic coefficients.
Journal of Differential Equations 253 (2012) 3113--3146.

\bibitem{BK14} A.Badanin,  E.Korotyaev.
Third order operator with periodic coefficients on the real axis.
St. Petersburg Math. J. 25:5 (2014) 713--734.

\bibitem{BK14x} A.Badanin, E.Korotyaev.
Sharp eigenvalue asymptotics for fourth order operators on the circle.
J. Math. Anal. Appl. 417 (2014), 804--818.

\bibitem{BK15} A.Badanin,  E.Korotyaev.
Trace formula for fourth order operators on unit interval. II.
Dynamics of PDE, Vol.12, No.3 (2015) 217--239.

\bibitem{BK21} A.Badanin, E.Korotyaev.
Third-order operators with three-point conditions associated
with Boussinesq's equation. Applicable Analysis (2021),
100(3), 527--560.



\bibitem{BK24x} A.Badanin, E.Korotyaev.
Mc'Kean's transformation for 3-rd order operators. ArXiv:2406.09668.

\bibitem{BK24xx} A.Badanin, E.Korotyaev.
Inverse problem for 3-rd order operators under the 3-point Dirichlet conditions.
ArXiv:2408.01886.

\bibitem{BK24xxx} A.Badanin, E.Korotyaev.
Inverse problem for the L-operator of the Lax pair
for the Boussinesq equation on the circle. (Russian)
Funct. Analysis and Its Appl., 58:3 (2024), 140--144,
transl. in arXiv:2408.01873.

\bibitem{BK24xxxx} A.Badanin, E.Korotyaev.
Asymptotics for 3-rd order operators on the line
with periodic small coefficients. To appear.



\bibitem{BBK06} A. Badanin, J. Br\"uning, E. Korotyaev.
The Lyapunov function for Schr\"odinger operators with
a periodic $2\times 2$ matrix potential.
J. Funct. Anal.  234 (2006) 106--126.




\bibitem{C00} R.Carlson. Compactness of Floquet isospectral sets for
the matrix Hill's equation. Proc. Amer. Math. Soc. 128 (2000),
 no. 10, 2933--2941

 \bibitem{CK06q} D. Chelkak; E. Korotyaev, Parametrization of the isospectral set
for the vector-valued Sturm-Liouville problem, J. Funct. Anal.
241(2006), 359--373.

\bibitem{CK09} D. Chelkak; E. Korotyaev, Weyl-Titchmarsh functions of
vector-valued Sturm-Liouville operators on the unit interval, Journal
Func. Anal., 257 (2009), 1546--1588.



\bibitem{CHGL00} S.Clark, H.Holden, F.Gesztesy, B.Levitan.
Borg-type theorem for matrix-valued Schr\"odinger and Dirac
operators. J. Diff. Eqs. 167(2000), 181-210

\bibitem{CK06}
 D.Chelkak,   E.Korotyaev.  Spectral estimates for Schr\"odinger
operator with periodic matrix potentials on the real line.
Int. Math. Res. Not., (2006), Art. ID 60314, 41 pp.

\bibitem{G50} I.M.Gelfand.
Expansion in characteristic functions
of an equation with periodic coefficients.
Proceedings of the USSR Academy of Sciences, 73 (1950), 1117--1120.

\bibitem{GL55} I.M. Gel'fand, V.B. Lidskii.
On the structure of regions of stability of linear canonical systems
of differential equations with periodic coefficients.
Usp. Mat. Nauk 10(1) (1955) 3--40, transl. AMS 2(8) (1958) 143--181.




\bibitem{HJ85} R.A. Horn, C.R. Johnson. Matrix Analysis.
Cambridge University Press, 1985.

\bibitem{KL77} V.K.Kalantarov,  O.A.Ladyzhenskaja.
Formation of collapses in quasilinear equations of parabolic and hyperbolic types.
(Russian) Boundary value problems of mathematical physics and related questions
in the theory of functions, 10.
Zap. Nauch. Sem. Leningrad. Otdel. Mat. Inst. Steklov.
(LOMI) 69 (1977), 77--102, 274.

\bibitem{K97}  E.Korotyaev.  The inverse problem for the Hill operator.
I. Int. Math. Res. Not. 1997, no. 3, 113--125.

\bibitem{K99}  E.Korotyaev.  Inverse Problem and
the trace formula for the Hill Operator, II.
Mathematische Zeitschrift 231(2) (1999) 345--368.

\bibitem{K08}  E.Korotyaev.
Spectral estimates for matrix-valued periodic Dirac operators.
Asymptot. Anal., 59 (2008), 195--225

\bibitem{K10}
E.Korotyaev.   Conformal spectral theory for the monodromy matrix.
Trans. Amer. Math. Soc.,  362 (2010), 3435--3462.

\bibitem{Kr55} M.G.Krein. The basic propositions of the theory
of $\l$-zones of stability of a canonical system of linear
differential equations with periodic coefficients.
In memory of A. A. Andronov,
Ed. Academy of Sciences of the USSR, 1955, 413--498;
transl. In Topics in Differential
and Integral Equations and Operator Theory (pp. 1--105), (1983),
Birkh\"auser, Basel.

\bibitem{MO75} V.A.Marchenko,  I.V.Ostrovskii.
Characteristics of the spectrum of the Hill operator. (Russian) Mat.
USSR Sb., 26 (1975), no.~4, 493--554.

\bibitem{McK81} H.McKean. Boussinesq's equation on the circle.
Com. Pure and Appl. Math. 34(1981) 599--691.

\bibitem{MM04} V.A.Mikhailets,  V.M.Molyboga.
Singular eigenvalue problems on the circle.
Methods Funct. Anal. Topology, 10(3) (2004), 44--53.

\bibitem{N67} M. A. Naimark.
Linear differential operators. Burns \& Oates, 1967.

\bibitem{P95}  V.Papanicolaou.
The spectral theory of the vibrating periodic beam.
Commun. Math. Phys. 170 (1995) 359 -- 373.

\bibitem{P03}  V.Papanicolaou. The Periodic Euler-Bernoulli
Equation. Transactions of the American Mathematical Society 355(9) (2003),
3727--3759.


\bibitem{PT87} J.P\"oschel, E.Trubowitz. Inverse spectral
theory. Academic Press, Boston, 1987.


\end{thebibliography}
\end{document}